\documentclass[journal,onecolumn,12pt,draftclsnofoot]{IEEEtran}
\IEEEoverridecommandlockouts
\usepackage{cite}
\usepackage{amsmath,amssymb,amsfonts,amsthm}

\usepackage[utf8]{inputenc} 
\usepackage[T1]{fontenc}

\usepackage{graphicx,dsfont}
\usepackage{textcomp}
\usepackage{xcolor,soul}
\usepackage{stmaryrd}
\usepackage{algorithm}
\usepackage{algpseudocode}
\usepackage{amsmath}
\usepackage{mathrsfs}
\usepackage{bm}
\usepackage[abs]{overpic}
\usepackage{dsfont}

\usepackage{physics}
\usepackage{tikz}
\usepackage{mathdots}
\usepackage{yhmath}
\usepackage{cancel}
\usepackage{siunitx}
\usepackage{array}
\usepackage{multirow}
\usepackage{tabularx}
\usepackage{extarrows}
\usepackage{booktabs}
\usetikzlibrary{fadings}
\usetikzlibrary{patterns}
\usetikzlibrary{shadows.blur}
\usetikzlibrary{shapes}

\theoremstyle{plain}
\newtheorem*{theorem*}{Theorem}
\newtheorem{theorem}{Theorem}

\newtheorem{lemma}{Lemma}

\newtheorem*{eg}{Example}

\usepackage{xcolor,varwidth}

\usepackage[scaled=.8]{beramono}
\newcommand{\bpara}[1]		{\medskip \noindent {\bf #1}}

\renewcommand\geq\geqslant
\renewcommand\leq\leqslant

\def\DR					    {{\textsf{DR}}}
\def\DE					{\stackrel{\rm{def}}{=}}
\newcommand\rob[1]			{\left( #1 \right)}
\newcommand\mat[1]		{\mathbf{#1}}
\newcommand\sqb[1]			{\left[ #1 \right]}
\newcommand\ym[2]			{#1_{\lambda}\sqb{#2}}

\newcommand\fig[1]			{Fig.~\ref{#1}}

\def\qo                     {q_0}

\def\eg					    {\emph{e.g.}~}
\def\ie					    {\emph{i.e.}~}

\def\e					    {{e}}

\usepackage{xspace}
\def\usalg					{{\fontsize{11pt}{11pt}\selectfont\texttt{US}-\texttt{Alg}}\xspace}

\def\lmimo					{$\lambda$--MIMO\xspace}
\def\mmimo					{M--MIMO\xspace}
\def\madc					{{$\mathscr{M}_\lambda$--{\fontsize{11pt}{11pt}\selectfont\texttt{ADC}}}\xspace}
\def\nbsc					{{\fontsize{11pt}{11pt}\selectfont\texttt{NBSC}}\xspace}
\def\mimoo				{{MIMO}--{OFDM}\xspace}

\def\madcs					{{$\mathscr{M}_\lambda$--{\fontsize{11pt}{11pt}\selectfont\texttt{ADCs}}}\xspace}
\newcommand{\iset}[1]       {\lfloor{\mathbb{#1}}\rceil}

\DeclareDocumentCommand{\sd}{o}  
{{\underline{\ast}\IfValueT{#1}{_{#1}}}}

\DeclareDocumentCommand{\xmod}{m o o o}  
{%
\IfNoValueTF{#4}
{{#1}\IfValueT{#2}{_{m,\mathsf{#2}}}\IfValueT{#3}{#3}}
{{#1}\IfValueT{#2}{_{m,\mathsf{#2}}^{\mathsf{#4}}}\IfValueT{#3}{#3}}
}

\newcommand{\MO}[1]		{\mathscr{M}_\lambda ({#1} )}
\newcommand{\VO}[1]		{\varepsilon_{#1}}
\newcommand{\normT}[3]{ {\| {#1} \|}_{ {{L}_{#2}} \left(  #3\right)}}
\newcommand{\normt}[3]{ {\| {#1} \|}_{ {{\ell}_{#2}} \left(  #3\right)}}
\newcommand{\BL}[1] 		{\mathcal{B}_{\Omega}}
\def\l{\left(}
\def\r{\right)}

\usepackage[inline]{enumitem}
\DeclareMathAlphabet{\mathsfit}{T1}{\sfdefault}{\mddefault}{\sldefault}
\SetMathAlphabet{\mathsfit}{bold}{T1}{\sfdefault}{\bfdefault}{\sldefault}
\usepackage{siunitx}

\def\BibTeX{{\mathrm B\kern-.05em{\sc i\kern-.025em b}\kern-.08em
    T\kern-.1667em\lower.7ex\hbox{E}\kern-.125emX}}
\begin{document}

\title{$\lambda$--MIMO: \\ Massive MIMO via Modulo Sampling
}

\author{Ziang Liu,
        Ayush Bhandari,
        and~Bruno Clerckx,~\IEEEmembership{Fellow,~IEEE}        
\thanks{The authors are with the Communications \& Signal Processing (CSP) Group at the Dept. of Electrical and Electronic Engg., Imperial College London, SW7 2AZ, UK. (e-mails:\{Ziang.Liu20,A.Bhandari,B.clerckx\}@imperial.ac.uk).}}

\maketitle

\begin{abstract}
Massive multiple-input multiple-output (M-MIMO) architecture is the workhorse of modern communication systems. Currently, two fundamental bottlenecks, namely, power consumption and receiver saturation, limit the full potential achievement of this technology. These bottlenecks are intricately linked with the analog-to-digital converter (ADC) used in each radio frequency (RF) chain. The power consumption in \mmimo systems grows exponentially with the ADC's bit budget while ADC saturation causes permanent loss of information. This motivates the need for a solution that can simultaneously tackle the above-mentioned bottlenecks while offering advantages over existing alternatives such as low-resolution ADCs. Taking a radically different approach to this problem, we propose \lmimo architecture which uses modulo ADCs (\madc) instead of a conventional ADC. Our work is inspired by the Unlimited Sampling Framework. \madc in the RF chain folds high dynamic range signals into low dynamic range modulo samples, thus alleviating the ADC saturation problem. At the same time, digitization of modulo signal results in high resolution quantization. In the novel \lmimo context, we discuss baseband signal reconstruction, detection and uplink achievable sum-rate performance. The key takeaways of our work include, (a) leveraging higher signal-to-quantization noise ratio (SQNR), (b)  detection and average uplink sum-rate performances comparable to a conventional, infinite-resolution ADC when using a $1$-$2$ bit \madc. This enables higher order modulation schemes \eg $1024$ QAM that seemed previously impossible, (c) superior trade-off between energy efficiency and bit budget, thus resulting in higher power efficiency. Numerical simulations and modulo ADC based hardware experiments corroborate our theory and reinforce the clear benefits of \lmimo approach. 
\end{abstract}


\newpage

\tableofcontents

\newpage

\section{Introduction}
\begin{figure}[tb]
	\begin{center}
		\includegraphics[width =0.65\textwidth]{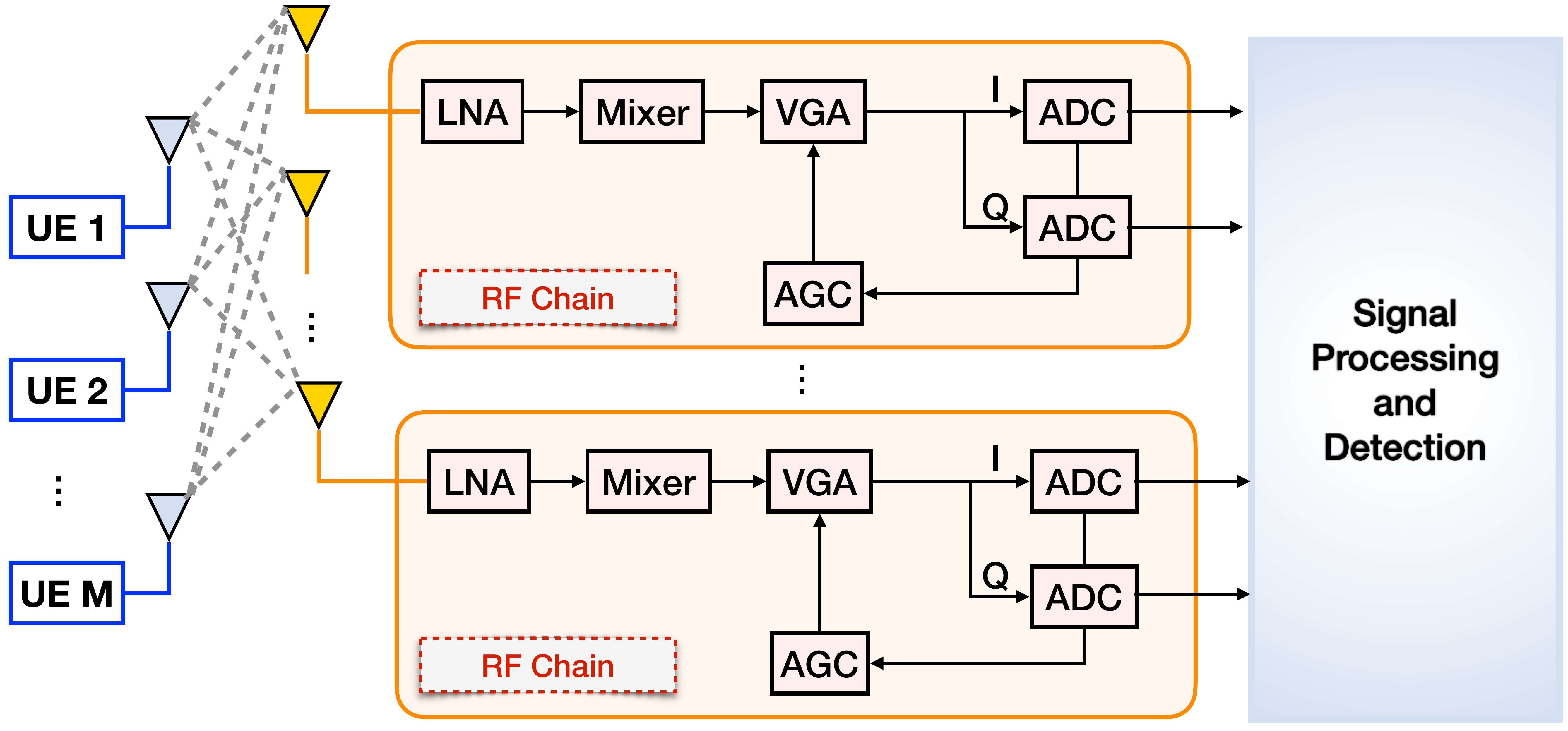}
		\caption{The receiver architecture of conventional massive MIMO (\mmimo) system with uplink user equipment (UE). Each RF chain consists of the LNA, mixer, VGA, AGC, and ADC.}
		\label{fig:rfchain}
	\end{center}
\end{figure}

As the name suggests, \emph{massive} multiple-input multiple-output (MIMO) or \mmimo architecture relies on a large number of antennas at the base station (BS). This allows for serving a large number of users. \mmimo has attracted much interest in the past decade because of its significant potential for \emph{spectral efficiency gain} \cite{larsson2014massive,andrews2014will}. However, the technological promise of \mmimo is limited by two fundamental challenges \cite{gao2016energy}, namely, 
\begin{enumerate*}
  \item \emph{power consumption}, and
  \item \emph{receiver saturation}.
\end{enumerate*}
It is established wisdom that these two bottlenecks inhibit the practical implementation of \mmimo in future wireless communication system. Specifically, each antenna in \mmimo is connected to a corresponding radio frequency (RF) chain consisting of 
\begin{enumerate*}[label=\roman*)]
  \item a low noise amplifier (LNA),
  \item a mixer,
  \item a variable gain amplifier (VGA) with an automatic gain controller (AGC), and 
  \item an analog to digital converter (ADC).
\end{enumerate*}
The RF-chain is illustrated in \fig{fig:rfchain}. Next, we elaborate on the two challenges that limit a \mmimo systems. 

\bpara{Challenge \#1: Power Consumption.} Among all the components in the RF chain, the ADC maps the received analog signal into the digital baseband samples and dominates the total power consumption of the RF-chain \cite{zhang2016spectral}. Note that the power consumption of the ADC grows linearly with the sampling rate but exponentially with the number of bits $b$ that decide the ADC's resolution. Thus, the total power consumption of \mmimo is a function of the ADC's bit budget. For example, the power consumption of the BS equipped with $256$ antenna elements, where each RF chain is equipped with two ADCs (quadrature pair), will be as high as \SI{256}{\watt} \cite{zhang2018low}, attributed to high bit resolution ($8$-$12$ bits) and high sampling rate ($\geq 20$ GSample/s). Such a power consumption is \emph{unsustainable} in realistic deployments \cite{zhang2018low}.

\begin{figure}[tb]
	\begin{center}
		\includegraphics[width = 1\textwidth]{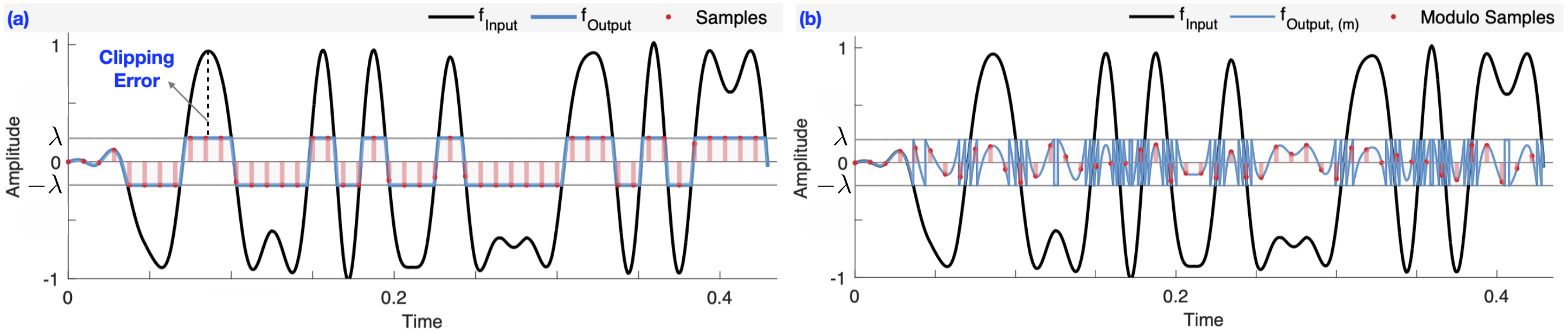}
		\caption{Illustration of output samples by conventional and \madcs. $f_\mathsf{Input}$: continuous input signal before ADC. $f_\mathsf{Output}$: continuous output signal shaped by limited ADC dynamic range. Samples: Sampled $f_\mathsf{Output}$ signals. a) In conventional ADC, saturation and clipping errors happen in $f_\mathsf{Output}$ signal and samples, because some values in $f_\mathsf{Input}$ exceeds dynamic threshold $\lambda$, b) In \madc, the input continuous signal $f_\mathsf{Input}$ is modulo-folded to $f_\mathsf{Output,(m)}$. No saturation occurs in the \madc.}
		\label{fig:sat}
	\end{center}
\end{figure}

\bpara{Challenge \#2: Receiver Saturation.} Another impediment to the \mmimo systems is due to receiver-ADC saturation; typically this is the case with wireless communication systems. This impediment is caused by the fact that ADCs are limited by their dynamic range (\DR), say $[-\lambda, \lambda]$ where the \emph{lower} threshold, $-\lambda$ and \emph{upper} threshold $\lambda$ of the ADC are assumed to be symmetric (as illustrated in \fig{fig:sat} (a)). When the received signal amplitude is larger than $\lambda$, the ADC output is distorted. This results in a clipped or saturated signal. Such undesirable yet unavoidable effects lead to high-frequency components in the digital signal, thus creating aliasing in the measurements. Consequently, the performance of the whole system  deteriorates.

\bpara{Previous Approaches for \mmimo.} To overcome the above stated problems---namely, \emph{high power consumption} and \emph{ADC saturation}---researchers typically resort to a solution strategy that disentangles the two challenges.

\begin{enumerate}[leftmargin = *,label = $\bullet$]
  \item High power consumption problem can be tacked by reducing the sampling rate and/or the bit resolution. Since Shannon's sampling theorem \cite{shannon1984communication} places a fundamental lower bound on the sampling rate, the only degree-of-freedom we can leverage is the bit resolution. This has been the driving force behind the conceptualization of  low-resolution ADC based solutions \cite{zhang2016spectral,zhang2018low,fan2015uplink,jacobsson2017throughput,choi2020advanced,mollen2016uplink}.
That said, low-resolution ADCs in \mmimo face a challenging signal detection problem due to the quantization noise and the non-linearity caused by the low-resolution quantization, resulting in sub-optimal system performance. One popular solution is to adopt low-order modulation schemes (\eg BPSK) to increase the inter-symbol distance in the constellation diagram, thus simplifying the signal detection problem. However, if a higher symbol rate is required, higher-order modulations will have to be adopted. As a result, signal detection becomes difficult and the system performance deteriorates. To improve the detection accuracy of higher-order modulations with low-resolution ADCs, two techniques are adopted in general, 
\begin{enumerate}[leftmargin = *,label = ---$\bullet$]
  \item \emph{Spatial Oversampling} \cite{jacobsson2017throughput,7247358} leverages multiple receive antennas receiving the same signal to improve the detection performance of high-order modulation (\eg 16 QAM).
  \item \emph{Temporal Oversampling} \cite{halsig2014information} leverages a higher sampling rate to increase the resolution of the ADC. 
\end{enumerate}

It is clear that both of these strategies entail an increased power consumption thus leading to an inherent trade-off between \emph{detection performance} and the \emph{power consumption}. 

\item To cope with the saturation problem, \emph{automatic gain control} (AGC) and \emph{variable gain amplifiers} (VGA) are adopted along with the ADC (cf.~\fig{fig:rfchain}) to ensure that the amplitude of the input signal is within the dynamic range. The AGC tracks the input signal and returns the feedback information of the signal amplitude range to the VGA. Then, the VGA adjusts the amplification of the input signal based on the feedback. Unfortunately, the received communication signal is fast-changing and not deterministic; thus the AGC may sometimes fail to track the range of the signal. Consequently, clipping errors still occur. Alternatively, the logarithmic ADC \cite{sit2004micropower,lee20092} can be used to extend the dynamic range of the ADC; however, the quantization resolution in the high amplitude region is coarse due to the logarithm operation, and information in this region may be lost, leading to system performance deterioration. 
\end{enumerate}

It is intuitive to note that strategies that take a decoupled approach, \ie address the high power consumption and receiver saturation problems independently, run into balancing trade-offs that induce limitations on the overall system performance. To the best of our knowledge, there is no combined method with satisfactory performance that solves these two problems in a holistic fashion. This motivates investigation of new approaches and we take a step in this direction by introducing modulo ADCs in the pipeline.

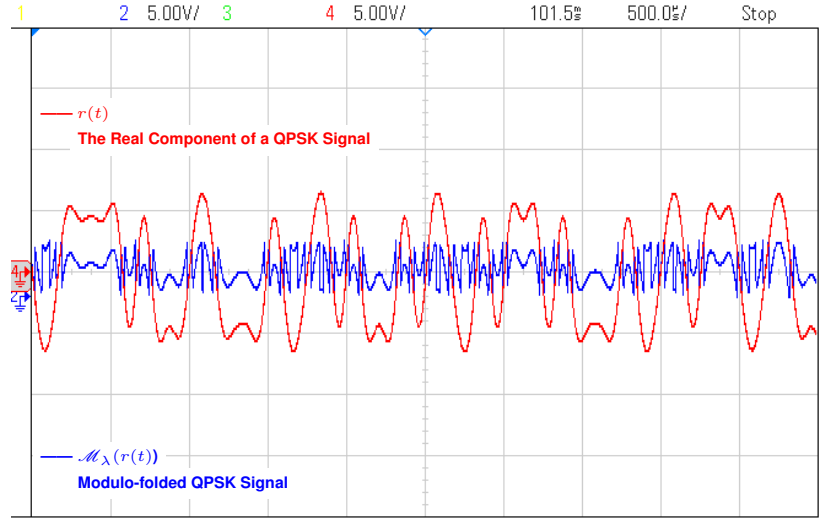
\begin{figure}[!t]
\centering
    \begin{overpic}[width=0.65\columnwidth]{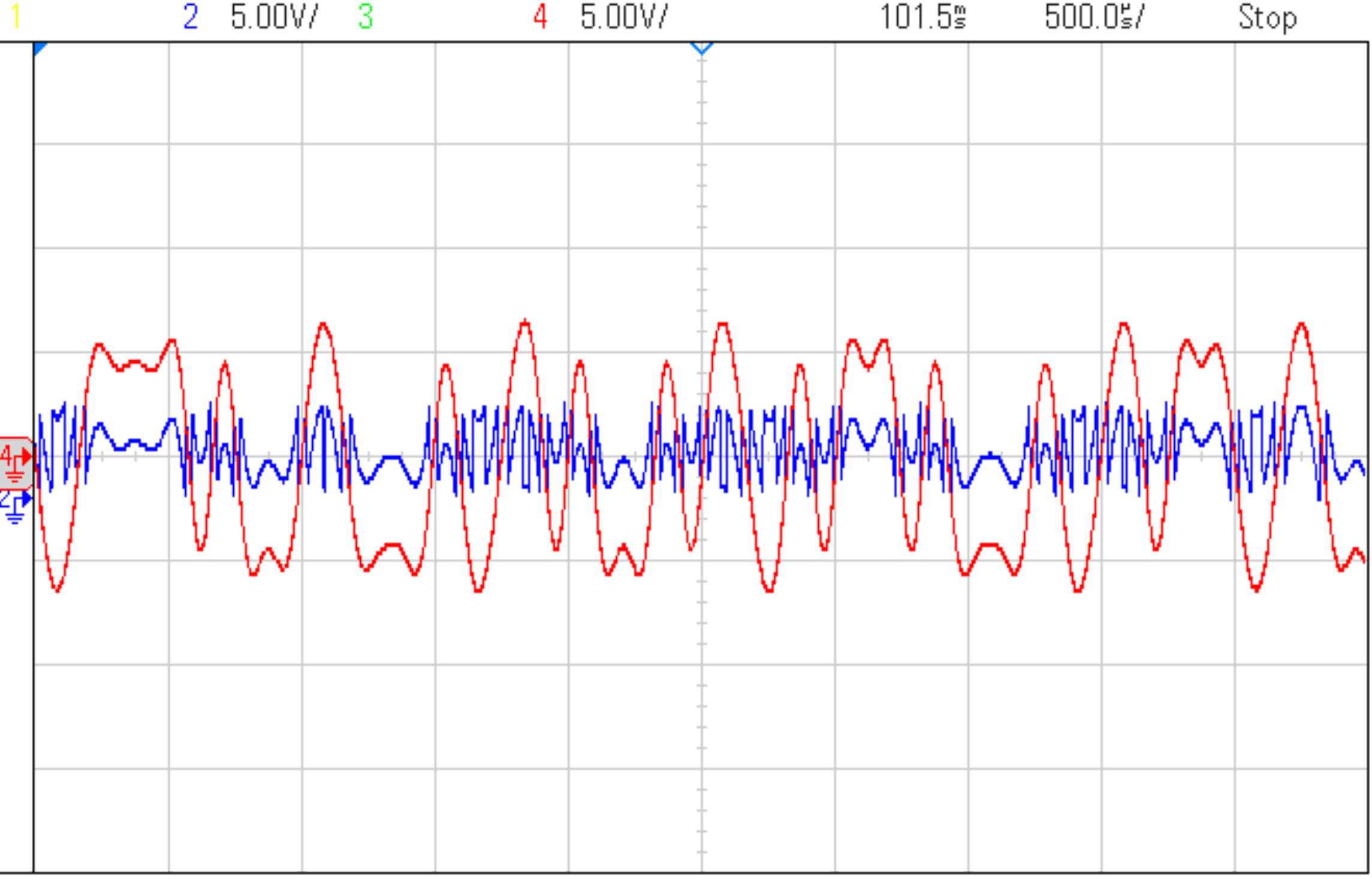}

    \put (8,152) {\colorbox{white}{\sf{\bfseries\fontsize{6pt}{6.5pt}\selectfont {\color{red}------ $r(t)$}}}}
    \put (22,142) {\colorbox{white}{\sf{\bfseries\fontsize{6pt}{6.5pt}\selectfont {\color{red} The Real Component of a QPSK Signal}}}}

    \put (8,22) {\colorbox{white}{\sf{\bfseries\fontsize{6pt}{6.5pt}\selectfont {\color{blue}------ $\mathscr{M}_{\lambda}({r(t)}$)}}}}
    \put (22,12) {\colorbox{white}{\sf{\bfseries\fontsize{6pt}{6.5pt}\selectfont {\color{blue} Modulo-folded QPSK Signal}}}}

    \end{overpic}
    \caption{Hardware output of \madc (cf.~\fig{fig:hd}) used in the Unlimited Sensing Framework \cite{Bhandari:2017:C,Bhandari:2020:Pata,Bhandari:2020:Ja,Bhandari:2021:J}. An oscilloscope screen shot plotting the real component of a quadrature phase shift keying (QPSK) signal (ground truth, red) and its folded version (blue). The input signal with dynamic range $13.6$V peak-to-peak ($\approx 7 \lambda$) is folded into a $4.02$V peak-to-peak signal.}
    \label{fig:scope_whole}
\end{figure}

\bpara{Our Approach: Unlimited Sampling for \mmimo.}
In this paper, we take a radically different approach to \mmimo and this is done by replacing a conventional ADC (cf.~\fig{fig:rfchain}) by a modulo ADC or \madc. The resulting approach is referred to as \lmimo. The \madcs are at the core of the {Unlimited Sensing Framework} (USF) \cite{Bhandari:2017:C,Bhandari:2020:Ja,Bhandari:2020:Pata, Bhandari:2018:Ca,Bhandari:2018:C,Bhandari:2019:C,Bhandari:2021:J,Bhandari:2022:J} which enables recovery of \emph{high dynamic range} (HDR) signals by acquiring \emph{low dynamic range} (LDR), modulo samples. The \madc maps a continuous-time signal into folded, digital samples by injecting a modulo non-linearity in the analog domain. This prevents distortion due to clipping/saturation that is prevalent in conventional ADCs \cite{Logan:1984,Abel:1991,Adler:2012,Esqueda:2016}. However, as shown in the hardware example in \fig{fig:scope_whole}, \madcs lead to a different form of information loss as one still needs to recover the original HDR signal from folded samples. This recovery is mathematically guaranteed provided that a \emph{Shannon-Nyquist like} density criterion is satisfied. We recall the modulo sampling theorem here. 

\begin{theorem}[Unlimited Sampling Theorem \cite{Bhandari:2017:C,Bhandari:2020:Ja}]
\label{thm:UST}
Let $f(t)$ be a continuous-time function with maximum frequency $\Omega$ (rads/s). Then, a sufficient condition for recovery of $f( t )$ from its modulo samples (up to an additive constant) taken every $T$ seconds apart is $T\leqslant 1/ \left( 2 \Omega \e\right)$ where $\e$ is Euler's number.
\end{theorem}
If this sufficient condition is satisfied, the recovery algorithms can be then used to recover the unfolded signal $f(t)$. Subsequently, the samples of the original HDR input are recovered by the unlimited sampling algorithm (\usalg) accompanying the theorem. A hardware example of modulo acquisition of a QPSK signal is shown in \fig{fig:scope_whole}. This is further used in Section~\ref{subsec: recovery} to provide a proof-of-concept demonstration. We refer to \fig{fig:sc_OFDM_recovery} and \fig{fig:hd} for a first look at our hardware experiments. Apart from \usalg, a ``Fourier-Prony'' method operating at much lower sampling rates was introduced in \cite{Bhandari:2021:J,Bhandari:2022:J}. Extensive experiments via a prototype \madc validated the empirical robustness of this approach while serving as the first hardware demonstration of USF. It was shown that signals as large as $\approx 25\lambda$ can be recovered via USF. With some side-information, \eg the knowledge of a certain number of unfolded samples, $f_\mathrm{s} > f_\mathsf{Nyquist}$ also suffices recovery \cite{Romanov:2019}, but uses a different recovery mechanism. Similarly, using multiple modulo thresholds \cite{Gong:2021:J} \eg Gaussian integers, allows for a different reconstruction approach. For other signal models and recovery methods, we refer to \cite{Rudresh:2018:C,Musa:2018:C,Ordentlich:2018:J,Bhandari:2020:C,FernandezMenduina:2021:J,Bhandari:2022:J,Florescu:2022:J}.

\begin{figure}[!t]
\begin{center}
\includegraphics[width =0.65\textwidth]{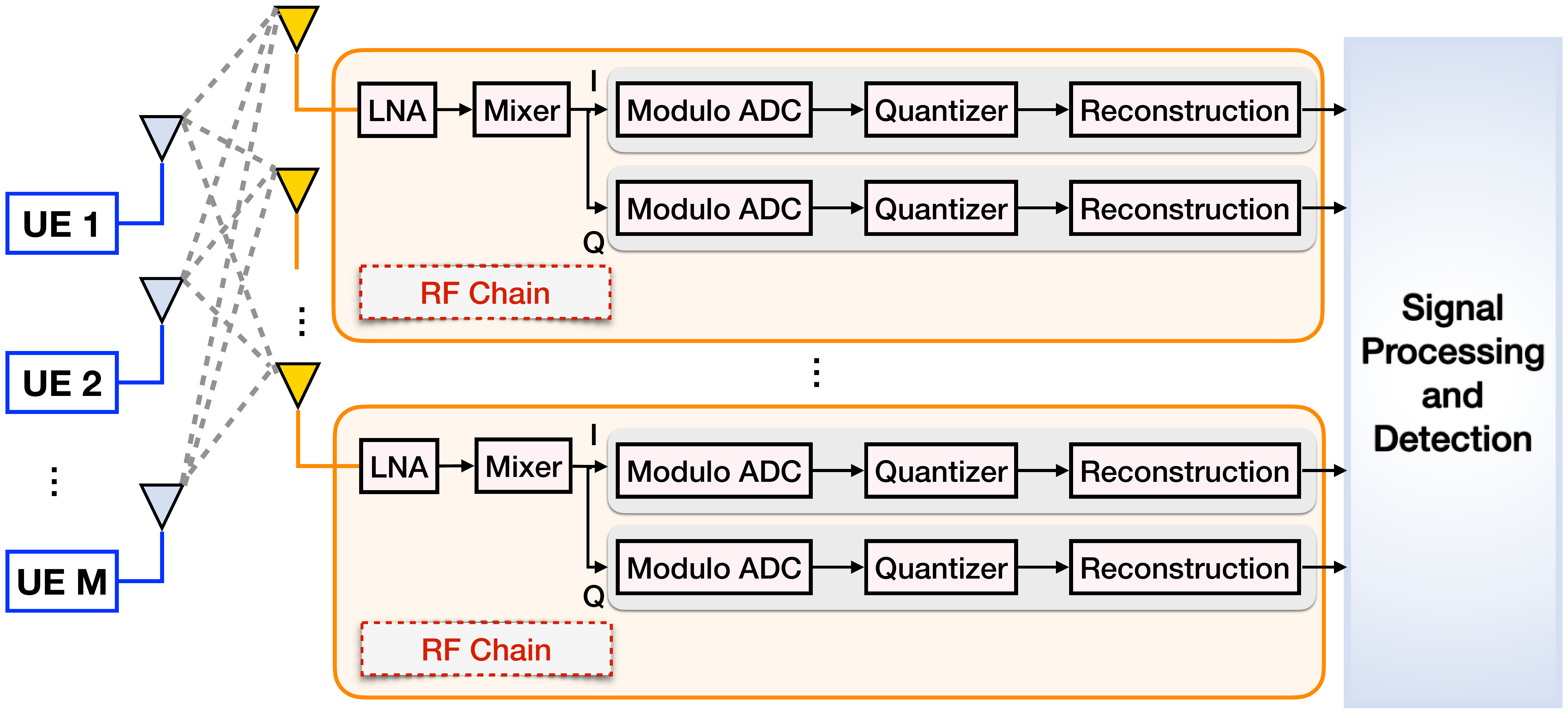}
\caption{Proposed \mmimo uplink system architecture with \madc in each RF chain.}
\label{fig:mimo}
\end{center}
\end{figure}

\bpara{Contributions and Overview of the Results.} We summarize our contributions as follows:
\begin{enumerate}[leftmargin = *, label = $\mathrm{C}_{\arabic*})$]
\item We formulate the USF acquisition and recovery problem in the context of 
two typical communication systems,
\begin{enumerate}[leftmargin = *,label =$\bullet$ ]
\item Narrowband single-carrier or the \nbsc case\footnote{Wideband single-carrier analysis is left for future work.}. 
\item Wideband, multi-carrier orthogonal frequency-division multiplexing (OFDM) \mmimo or \mimoo.
\end{enumerate}
For the two setups, we consider both \emph{reconstruction} and \emph{detection} performances. Here, our key contribution is the receiver design for USF enabled \mmimo. Since $\mathscr{M}_\lambda$ maps a bandlimited function into a non-bandlimited one, existing receiver architectures do not apply to our case and this is the non-trivial aspect of our work. Our receiver design enables $1024$ QAM for both (a) narrowband single-carrier and (b) wideband OFDM signal and detection is possible with as little as 2 bit resolution; this has not been reported in previous literature. 
\item We present the \emph{Achievable sum-rate Analysis} for our new approach considering different bit resolutions. Surprisingly, with few bits \eg $1$- or $2$-bits, the \lmimo architecture achieves a similar sum-rate performance when compared with $\infty$-bit resolution, conventional ADC. We attribute this finding to the fact that for a given bit budget, the \madc offers a higher quantization resolution, thus resulting in a lower quantization noise. We also provide \emph{approximates} for the achievable sum-rate in the context of,
\begin{enumerate}[leftmargin = *]
    \item Maximum ratio combiner.
    \item Zero forcing combiner. 
\end{enumerate}
\item We analyze the energy efficiency of the \lmimo approach and contrast it with low-resolution, conventional ADC as a baseline. Numerical results illustrate that the energy efficiency of the \madc is higher than the conventional ADC while offering competitive detection performance (\eg $1024$ QAM), as listed in $\mathrm{C}_{1})$.
\item Finally, we validate our approach via computer simulations as well as experiments based on the \madc hardware \cite{Bhandari:2021:J}, thus corroborating our theoretical findings.    
\end{enumerate}

\bpara{Key Benefits of our Approach.} We study and demonstrate the benefits of the \lmimo approach. The benefits are pivoted on the fact that \madcs, when coupled with a tailored receiver design, allow for simultaneously solving the power consumption and receiver saturation problems of conventional \mmimo communications. Our new approach offers several benefits:
\begin{enumerate}[leftmargin = *, label = \arabic*)]
\item Our energy efficiency analysis shows that for a given bit budget \lmimo approach achieves \emph{lower power consumption}.

\item Our approach offers similar \emph{detection} and \emph{sum-rate performances} as an $\infty$-resolution, conventional ADC \mmimo which remains unachievable with other methods and architectures proposed in the literature. Satisfactory detection and sum-rate performances are realized in the \lmimo setup as we fundamentally avoid the ADC saturation problem (cf.~\fig{fig:sat} (b)). At the same time, our approach entails a \emph{lower} signal-to-quantization-noise ratio (SQNR); for the same bit budget, lower dynamic range of the modulo ADCs implies higher resolution. This is elaborated in Section-\ref{subsec:metric}.

\item Our approach is devoid of VGA and AGC apparatus that is typically used to cope up with ADC saturation problem, thus enabling \emph{lower complexity} hardware.
\end{enumerate}

Although, recently an attempt has been made to address the receiver saturation through the modulo folding operation in a full-duplex setup \cite{ordonez2021full}, this study does not consider signal recovery and detection aspects.  In comparison, our receiver design considers the full-stack of signal processing operations, namely reconstruction of the modulo-folded signals and signal detection, as well as the achievable sum-rate analysis.

\bpara{Organization of this Paper.} The rest of the paper is organized as follows. We revisit the system models for multi-user (MU) \mmimo signals in Section \ref{sec:systemmodel}. We introduce the receiver design with \madc in Section \ref{sec:moduloadc}. In Section \ref{sec:asr}, we analyze the SQNR and achievable uplink sum-rate when incorporating the \madc in the narrowband setup. Numerical and hardware experiments are presented in  Section \ref{sec:result}. Finally, we conclude this work in Section \ref{sec:con}.

\bpara{Notation.} The set of natural numbers, reals, integers, and complex numbers are denoted by $\mathbb{N}$, $\mathbb{R}$, $\mathbb{Z}$ and $\mathbb{C}$, respectively. For $K \in \mathbb{R}$, we use $\iset{K}$ to denote the non-negative integers until $K$, that is, $\{k: k \in \mathbb{Z} \cap[1, K+1)\}$. For $x \in \mathbb{R}$, we define the floor and the ceiling operations by $\lfloor x \rfloor = \sup \{k \in \mathbb{Z}: k \leqslant x\}$ and $\lceil x\rceil=\inf \{k \in \mathbb{Z}: k \geqslant x\}$. We use $\langle , \rangle$ to denote the inner-product defined by $\langle a, b \rangle = \int a(x) b^*(x) dx$. The $L^\textrm{th}$ order finite difference of the sequence $x[k]$ is defined by the recursion, $\left(\Delta^{L} x\right)[k]=\Delta^{L-1}(\Delta x)[k], L > 1$. When $L = 1$, $(\Delta x)[k] {=} x[k+1]-x[k]$. Continuous signals and discrete sequences are expressed as $x(t), t \in \mathbb{R}$ and $x[k], k \in \mathbb{Z}$, respectively. The semi-discrete convolution operator $\sd[T]$ with sampling interval $T$ is defined as the interpolation map 
\[
\sd[T]: c \in \ell_2 \longmapsto
\rob{c \sd[T] \phi}\rob{t}
=\sum\nolimits_{n\in\mathbb{Z}}c\sqb{n} \phi\rob{t-nT}.
\]
When $T=1$, we denote the above by $\sd$.
A complex number, say $x \in \mathbb{C}$, is represented as $x=\Re(x) + \jmath \Im(x)$. Matrices, vectors and scalars are written in capital boldface, small boldface and normal fonts, \ie $\mathbf{X}$, $\mathbf{x}$ and $x$, respectively. $[\mathbf{X}]_{i,j}$ denotes the entry of the matrix at index $(i, j)$. Similarly, $[\mathbf{x}]_{i}$ for vector $\mathbf{x}$. We use $\mathbf{X}^H$ and $\mathbf{X}^\top$ to denote the conjugate-transpose and transpose operations, respectively, of the matrix $\mathbf{X}$. We use $\mathscr{E}{(\cdot)}$, $|\cdot|$, and $\|\cdot\|_{\ell_p}$ to denote statistical expectation, absolute value and Euclidean norm. When $p = \infty$, the norm denotes the max-norm, or $\|\cdot\|_{\infty}$. We say $f \in C^{N}(\mathbb{R})$ is $\Omega$-bandlimited or, $f \in \mathcal{B}_{\Omega} \Leftrightarrow \widehat{f}(\omega)=\mathds{1}_{[-\Omega, \Omega]}(\omega) \widehat{f}(\omega)$ where $\mathds{1}_{\mathcal{D}}(t)$ is the indicator function on domain $\mathcal{D}$.

\begin{figure}[tb]
\centering
\includegraphics[width =0.65\textwidth]{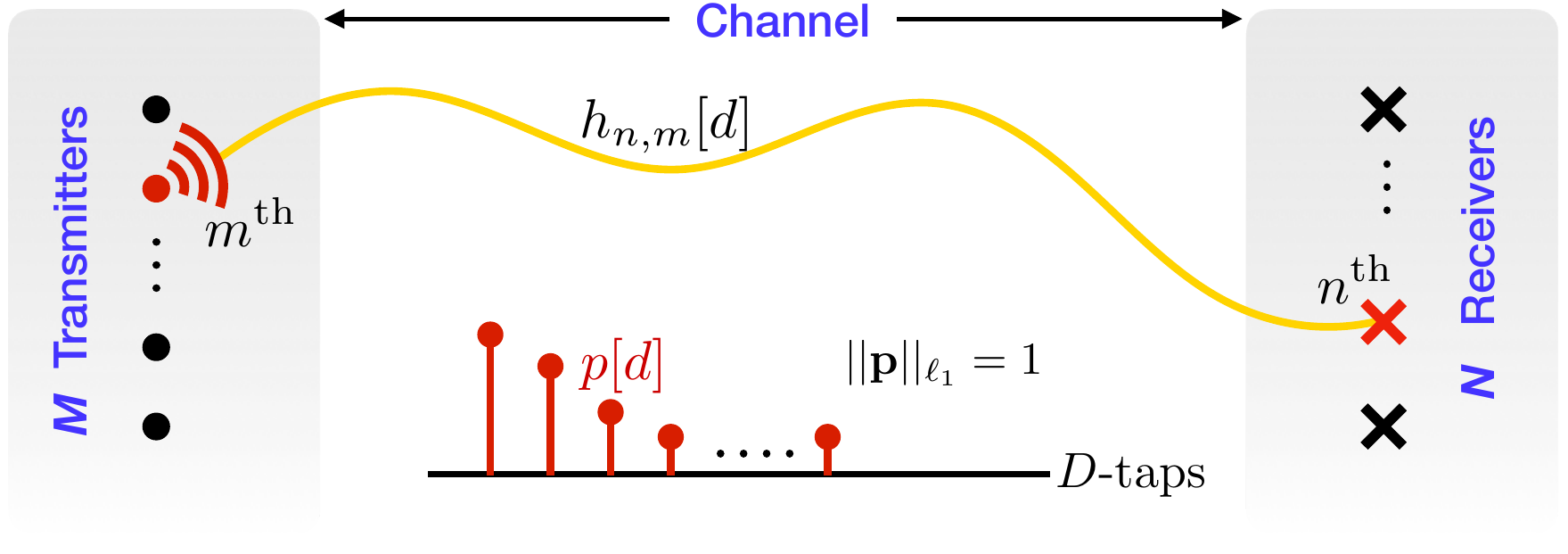}
\caption{Channel model between $m^\textrm{th}$ user and $n^\textrm{th}$ antenna in the BS.}
\label{fig:channel}
\end{figure}
\section{System model}
\label{sec:systemmodel}
In this section, we revisit the transmitter, channel, and receiver models, which will help us build the receiver architecture of the \lmimo system. As shown in Fig. \ref{fig:channel}, we assume that there are $M$ single-antenna users and one BS equipped with $N$ receive antenna elements (with $N \gg M$) in one cell. Note that the users transmit the information to the BS in the same time-frequency band. We denote $\iset{K}$, $\iset{M}$, $\iset{N}$, and $\iset{D}$ as the set of indices of subcarriers, users, receive antennas, and channel taps, respectively.

\bpara{Transmitter.} At each user location, we revisit the transmitter structure. To demonstrate the applicability of the \madc in communication systems, we consider two popular scenarios, namely, \nbsc and wideband \mimoo systems. 

\subsubsection{Narrowband Single-Carrier Signal Model (\nbsc)}
Let us begin our discussion with the definition of the baseband signal denoted by $\xmod{x}[b][\rob{t}]$. Starting with an upsampled sequence
$\xmod{x}[b][\sqb{k}][\uparrow]$, filtering the same with a pulse-shaping filter\footnote{The pulse-shaping filter can be the raised-cosine Nyquist pulse\cite{beaulieu2004parametric}, with excess bandwidth $\alpha, 0 \leq \alpha \leq 1$, given by
\begin{equation}\label{eq:nypulse}
f_{\mathsf{RC}}(t)=\operatorname{sinc}\rob{\frac{t}{ T_\mathsf{rep}}}
\frac{\cos (\pi \alpha t / T_\mathsf{rep})}{1-4 \alpha^{2} t^{2} / T_\mathsf{rep}^{2}},
\end{equation}
where $T_\mathsf{rep} = 1/(2\Omega)$ is the symbol repetition rate.} results in the continuous-time signal, 
\begin{equation}
\label{eq:NBSCb}
\underbrace{
\xmod{x}[b][\sqb{k}][\uparrow]
\longrightarrow
\boxed{\psi}}_{\textsf{Pulse-Shaping}}
\longrightarrow
\underbrace{
\xmod{x}[b][\rob{t}]   \DE \rob{\xmod{x}[b][\sqb{k}][\uparrow] \sd[T]\psi}\rob{t}}
_{\textsf{Baseband Signal}}.
\end{equation}
There on, given the same for $\mathsf{I}$/$\mathsf{Q}$ channels, denoted by $\xmod{x}[b][\rob{t}][I]$ and $\xmod{x}[b][\rob{t}][Q]$, respectively, we generate the passband signal by modulating the baseband sources, 
\begin{equation}
\begin{aligned}
    x_{m,\mathsf{p}}(t) & = 
   \left[ {\begin{array}{*{20}{c}}
  {\cos \left( {{\omega _c}t} \right)}&{} \\ 
  {}&{ - \sin \left( {{\omega _c}t} \right)} 
\end{array}} \right]\left[ {\begin{array}{*{20}{c}}
  {\xmod{x}[b][\rob{t}][Q]} \\ 
  {\xmod{x}[b][\rob{t}][I]}
\end{array}} \right] \equiv {{\mathbf{M}}_{{\omega _c}}}{{\mathbf{x}}_m}  = \sqrt{2} \Re\rob{\xmod{x}[b][\rob{t}] e^{\jmath \omega_{\mathrm{c}} t}},
\end{aligned}
\end{equation}
where $\mat{M}_{\omega_c}$ is the modulation matrix parameterized by the carrier frequency $\omega_\mathrm{c} = 2 \pi f_\mathrm{c}$ (rads/sec). 

\subsubsection{Wideband OFDM Signal Model (\mimoo)}
The baseband OFDM signal is denoted by $x_{m,\mathsf{b}}^\prime(t)$. Starting with a modulated complex sequence of symbols ${\widehat{{x}}}_{m, \mathsf{b}}\sqb{k}$ to be transmitted using OFDM. Its vector representation is $\widehat{\mathbf{x}}_{m, \mathsf{b}} = [{\widehat{{x}}}_{m, \mathsf{b}}\sqb{1}, \ldots, {\widehat{{x}}}_{m, \mathsf{b}}\sqb{K}]^\top$. By using the $K$-point inverse discrete Fourier transformation (IDFT), the sequence of the baseband OFDM symbol ${{x}}^\prime_{m, \mathsf{b}}[k]$ with vector expression ${\mathbf{x}}^\prime_{m, \mathsf{b}} = [{{{x}}}^\prime_{m, \mathsf{b}}\sqb{1}, \ldots, {{{x}}}^\prime_{m, \mathsf{b}}\sqb{K}]^\top$ is given by
\begin{equation}
    \mathbf{x}_{m,\mathsf{b}}^\prime =  \mathbf{F}^H_\mathsf{DFT} {\widehat{\mathbf{x}}}_{m, \mathsf{b}}, 
\end{equation}
where $\mathbf{F}_\mathsf{DFT} \in \mathbb{C}^{K \times K}$ is the normalized discrete Fourier transform (DFT) matrix with element,
\begin{equation}
    [\mathbf{F}_\mathsf{DFT}]_{i,j} = \frac{1}{\sqrt{K}} e^{\frac{-\jmath 2 \pi(i-1)(j-1)}{K}}.
    \label{eq:dft}
\end{equation}
There on, a cyclic prefix with length $N_\mathsf{cp}$ is prepended to the sequence $ x_{m,\mathsf{b}}^\prime[k]$ to compensate for the multipath effect, where $x_{m,\mathsf{b}}^\prime[k] = x_{m,\mathsf{b}}^\prime[K + k]$ for $k = -(N_\mathsf{cp}-1), ..., 0$. We then upsample ${{x}}^\prime_{m, \mathsf{b}}[k]$ resulting in the upsampled sequence $\xmod{x}[b][\sqb{k}][\prime\uparrow]$. Subsequently,  it is filtered with a pulse-shaping filter leading to the continuous-time OFDM signal, 
\begin{equation}
\label{eq:OFDMs}
\underbrace{
\xmod{x}[b][\sqb{k}][\prime\uparrow]
\longrightarrow
\boxed{\psi}}_{\textsf{Pulse-Shaping}}
\longrightarrow
\underbrace{
\xmod{x}[b][\rob{t}][\prime]   \DE \rob{\xmod{x}[b][\sqb{k}][\prime\uparrow] \sd[T]\psi}\rob{t}}
_{\textsf{Baseband OFDM Signal}}.
\end{equation}
Finally, we generate the passband signal by modulating the baseband $\mathsf{I}$/$\mathsf{Q}$ channels, denoted by $\xmod{x}[b][\rob{t}][\prime{I}]$ and $\xmod{x}[b][\rob{t}][\prime{Q}]$, respectively
\begin{equation}
\begin{aligned}
    x^\prime_{m,\mathsf{p}}(t) & = 
   \left[ {\begin{array}{*{20}{c}}
  {\cos \left( {{\omega _c}t} \right)}&{} \\ 
  {}&{ - \sin \left( {{\omega _c}t} \right)} 
\end{array}} \right]\left[ {\begin{array}{*{20}{c}}
  {\xmod{x}[b][\rob{t}][\prime{Q}]} \\ 
  {\xmod{x}[b][\rob{t}][\prime{I}]}
\end{array}} \right] \equiv {{\mathbf{M}}_{{\omega _c}}}{{\mathbf{x}}^\prime_m}  = \sqrt{2} \Re\rob{x^\prime_{m,\mathsf{b}}(t) e^{\jmath \omega_{\mathrm{c}} t}}.
\end{aligned}
\end{equation}
Therefore, we can generalize the passband signals for the two
\begin{equation}
x_m(t) \!  = \!   
\begin{cases}\sqrt{2} \Re\left(x_{{m,\mathsf{b}}}(t) e^{\jmath \omega_{\mathrm{c}} t}\right), &\! \!  \text {if  single-carrier } \\ 
\sqrt{2} \Re\left(x_{{m,\mathsf{b}}}^\prime(t) e^{\jmath \omega_{\mathrm{c}} t}\right), &\! \!  \text {if  OFDM }
\end{cases},
\end{equation}
which is transmitted at each user location into the wireless channel.

\bpara{Channel Model.} We revisit the wideband \mimoo model. It can be simplified to the narrowband channel model with single path.
The channel of the wideband \mimoo system encapsulates multipath propagation and hence, it is modeled by $D$ taps $(D \leq N_\mathsf{cp})$. As shown in \fig{fig:channel}, the $d^\textrm{th}$ path impulse response $h_{{n,m}}\sqb{d}$ between the $m^{\textrm{th}}$ user and the $n^{\textrm{th}}$ antenna element at the BS is the product of a large-scale fading coefficient $\sqrt{\eta_m}$ and a small-scale fading parameter $g_{{n,m}}[d]$ and is given by,
\begin{equation}
h_{{n,m}}[d]=\sqrt{\eta_m} \, g_{{n,m}}[d].
\label{eq:hmn}
\end{equation}
Here, we assume that the large fading coefficients are the same across the antenna array at the BS, therefore we only have one coefficient $\eta_m$ between the $m^{\textrm{th}}$ user and all antenna elements. 

Following the conventional propagation model \cite{tse2005fundamentals}, we assume that the channel fading follows an i.i.d.~Rayleigh distribution, or $g_{{n,m}}[d] \sim \mathcal{C N}(0, p[d])$. Hence, the variance of the small-scale fading in each path is its power delay profile, or  $\mathscr{E}[\left|g_{{n,m}}[d]\right|^{2}] = p[d]$ that satisfies the relation, 
\begin{equation}
\|\mathbf{p}\|_{\ell_1}=1, \quad
{\mathbf{p}} = {\left[ {\begin{array}{*{20}{c}}
  {p\left[ 1 \right]}& \cdots &{p\left[ D \right]} 
\end{array}} \right]^ \top }, \quad p\sqb{d}>0.
\end{equation}

In the \nbsc \mmimo system, since the multiple paths are not resolvable in narrow bandwidth, a single path (frequency-flat) model is used, thus $D$ reduces to $1$. The frequency-flat channel is expressed as $\mat{H} \in \mathbb{C}^{N \times M}$,
\[{\mathbf{H}} = \underbrace {\left[ {\begin{array}{*{20}{c}}
  {{g_{1,1}}}& \cdots &{{g_{1,N}}} \\ 
   \vdots & \ddots & \vdots  \\ 
  {{g_{M,1}}}& \cdots &{{g_{M,N}}} 
\end{array}} \right]}_{\mathbf{G}}\underbrace {\left[ {\begin{array}{*{20}{c}}
  {\sqrt {{\eta _1}} }&{}&{} \\ 
  {}& \ddots &{} \\ 
  {}&{}&{\sqrt {{\eta _M}} } 
\end{array}} \right]}_{{{\mathbf{D}}_{\boldsymbol\eta} }} \equiv 
\mat{G}\mat{D}_{\boldsymbol{\eta}},
\]
where $g_{{n,m}} \sim \mathcal{CN} \rob{0, 1}$. For both channels, the channel state information (CSI) is assumed to be known at the receiver end. 

\bpara{Receiver.} The received signal at the $n^\textrm{th}$ antenna is given by 
\begin{equation}
z_n(t) \DE \sum\limits_{m \in \iset{M}} \sqrt{p_\mathrm{u}}{\left( {{h_{n,m}}{\,{\underline  *  }_T\,}{x_m}} \right)} \left( t \right) + \epsilon_n(t),
\label{eq:ztime}
\end{equation}
where $p_\mathrm{u}$ is the average transmit power at the user, and $\epsilon_n(t) \sim \mathcal{C N}(0, 1), \forall n \in \iset{N}$ is the additive Gaussian noise at the receiver.

\begin{figure*}[t]
\centering
		\includegraphics[width = 0.8\textwidth]{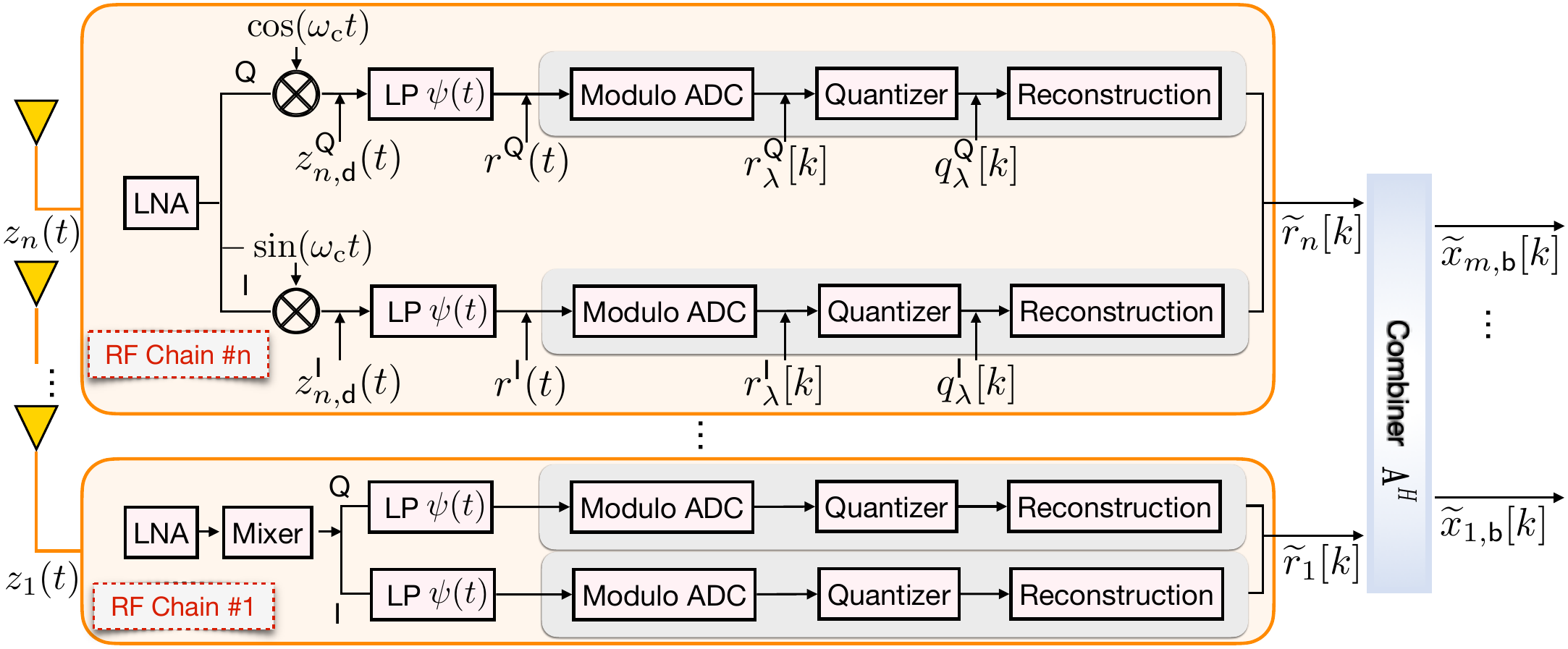}
		\caption{Receiver architectures with \madcs for single-carrier system.}
		\label{fig:flowa}
\end{figure*}

\begin{figure}[t]
	\begin{center}
		\includegraphics[width = 0.65\textwidth]{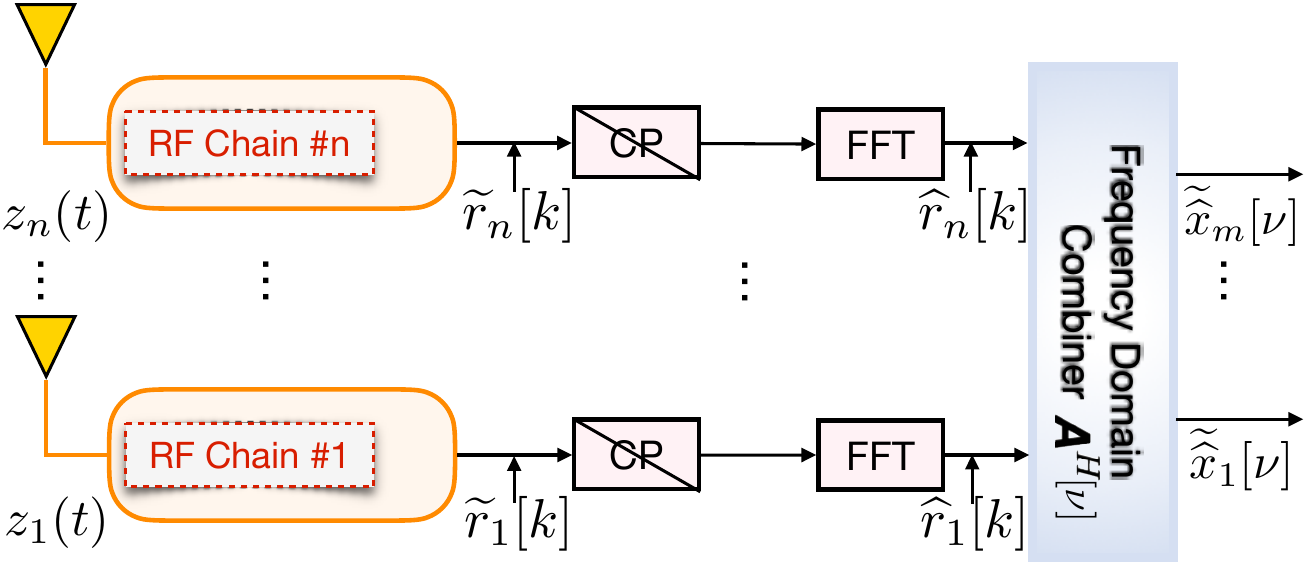}
		\caption{Receiver architectures with \madcs for multi-carrier (OFDM) system.}
		\label{fig:flowb}
	\end{center}
\end{figure}

 Subsequently, the received signal at each antenna enters the RF chain and is processed using the \madc receiver. The novel RF chains and signal processing blocks for both single-carrier and multi-carrier (OFDM) signals are shown in \fig{fig:flowa} and \ref{fig:flowb}, respectively.

 At the receiver side, the continuous-time signal $z_n(t)$ is firstly amplified by LNA and demodulated to the baseband by means of coherent detection. This is facilitated by, 
 \begin{equation}
\renewcommand\arraystretch{1.3}
\begin{bmatrix}
  z^\mathsf{Q}_{n,\mathsf{d}}(t) \\ z^\mathsf{I}_{n, \mathsf{d}}(t) 
\end{bmatrix}
=
\begin{bmatrix}
  \cos(\omega_\mathrm{c} t) &   \\  & -\sin(\omega_\mathrm{c} t)
\end{bmatrix}
\begin{bmatrix}
  z_n(t) \\ z_n(t)
\end{bmatrix} 
\equiv {{\mathbf{M}}_{{\omega _c}}}{{\mathbf{z}}_n}.
\end{equation}
Subsequently, the quadrature and in-phase signals are low-pass filtered with $\psi(t)$; this eliminates the images located outside from the baseband. Quantitatively, we have, 
\begin{equation}
r_n^\mathsf{Q}(t)  = \langle z^\mathsf{Q}_{n,\mathsf{d}}, \overline{\psi} \rangle \quad \mbox{and} \quad r_n^\mathsf{I}(t) = \langle z^\mathsf{I}_{n,\mathsf{d}}, \overline{\psi} \rangle,
\label{eq:down}
\end{equation}
where $\overline{\psi}(t) = \psi(-t)$ (time-reversal).
Then, the baseband signal is acquired by the \madc to reduce power consumption and prevent ADC saturation, as detailed in Section \ref{sec:moduloadc}. 
\subsubsection{Single-Carrier Signal Detection}
It is well known that maximum likelihood detector achieves the optimal performance in theory \cite{van1976maximum}. However, the computational complexity increases exponentially with user number $M$. Thus, a commonly adopted approach is to use the maximum ratio combining (MRC) or zero forcing (ZF) linear combiners \cite{clerckx2013mimo}. This is used to separate the streams from each uplink user. The matrix form of the combiner is given by
\begin{equation}
\widetilde{\mathbf{x}}_{\mathsf{b}} =\mathbf{A}^{H} \mathbf{r},
\end{equation}
where $\widetilde{\mathbf{x}}_{ \mathsf{b}} \in \mathbb{C}^{M \times 1}$ is the estimated baseband signal for all $M$ users, $\mathbf{r} \in \mathbb{C}^{N \times 1}$ is the recovered signal for all $N$ antennas, and $\mathbf{A}^H$ is the combiner where,
\begin{equation}
\mathbf{A}=\left\{\begin{array}{lc}
\mathbf{H}, & \text {\; for MRC } \\
 \mathbf{H} (\mathbf{H}^{H} \mathbf{H} )^{-1}, & \text {for ZF }.
\end{array}\right.
\label{eq:combiner}
\end{equation}

\subsubsection{OFDM Signal Detection}
As shown in \fig{fig:flowb}, we proceed by undoing the cyclic prefix (CP). Then, the samples at $n^\textrm{th}$ antenna are transformed to frequency domain by fast Fourier transform (FFT), which yields the frequency-domain signal $\widehat{\mathbf{r}}_n$
\begin{equation}
    \widehat{\mathbf{r}}_n = \mathbf{F}_\mathsf{DFT}\mathbf{r}_n.
\end{equation}
The received signal at $\nu^\textrm{th}$ subcarrier can also be written as \eqref{eq:yfreq}
\begin{equation}
[\widehat{\mathbf{r}}_n]_{\nu} \equiv \widehat{r}_n[\nu]=\sum_{m \in \iset{M}} \sqrt{ p_\mathrm{u} \eta_m } \, \widehat{g}_{n,m}[\nu] \widehat{x}_m[\nu]+\widehat{\epsilon}_n[\nu], \quad \forall n,
\label{eq:yfreq}
\end{equation}
where $\widehat{\mathbf{g}}_{n,m}\in \mathbb{C}^{K}, [\widehat{\mathbf{g}}_{n,m}]_\nu = \widehat{g}_{n,m}[\nu]$ is the frequency response of the small-scale fading between $n^\textrm{th}$ antenna and $m^\textrm{th}$ user, 
\begin{equation}
\label{eq:freqg}
    \widehat{\mathbf{g}}_{n,m} = \mathbf{F}_\mathsf{DFT} \mathbf{g}^{\mathsf{pad}}_{n,m}.
\end{equation}
In \eqref{eq:freqg}, $\mathbf{g}^{\mathsf{pad}}_{n,m} = [\mathbf{g}_{n,m}^\top \; | \; \mathbf{0}_{1 \times (K-D)}]^\top $, and $\widehat{\epsilon}_n[\nu] \sim \mathcal{C N}(0, 1)$ is the unitary Fourier transform of the noise vector. Therefore, the transmit-receive relation at the $\nu^\textrm{th}$ subcarrier can be written in matrix form as, 
\begin{equation}
    \widehat{\mathbf{r}}[\nu] = \sqrt{p_\mathrm{u}}\widehat{\mathbf{H}}[\nu]\widehat{\mathbf{x}}[\nu] + \widehat{\boldsymbol{\epsilon}}[\nu],
\end{equation}
where $[\widehat{\mathbf{H}}[\nu]]_{n,m} =  \eta_m \widehat{g}_{n,m}[\nu]$ is the frequency response of the channel at the $\nu^\textrm{th}$ subcarrier. Subsequently, frequency domain combiner given in \eqref{eq:combinerfreq} is adopted on each subcarrier, and the estimated data $\widetilde{\widehat{\mathbf{x}}}[\nu] = \widehat{\mathbf{A}}^H[\nu] \widehat{\mathbf{r}}[\nu]$ from uplink users at the $\nu^\textrm{th}$ subcarrier is obtained \cite{mollen2016uplink}, where
\begin{equation}\label{eq:combinerfreq}
\widehat{\mathbf{A}}[\nu]=\left\{\begin{array}{lc}
\widehat{\mathbf{H}}[\nu], & \text {\; for MRC } \\
\widehat{\mathbf{H}}[\nu] (\widehat{\mathbf{H}}^H[\nu] \widehat{\mathbf{H}}[\nu] )^{-1}, & \text {for ZF }.
\end{array}\right.
\quad \forall \nu,
\end{equation}
\section{Receiver Design with \madc}
\label{sec:moduloadc}
\begin{figure}[tb]
	\begin{center}
		\includegraphics[width =0.65\textwidth]{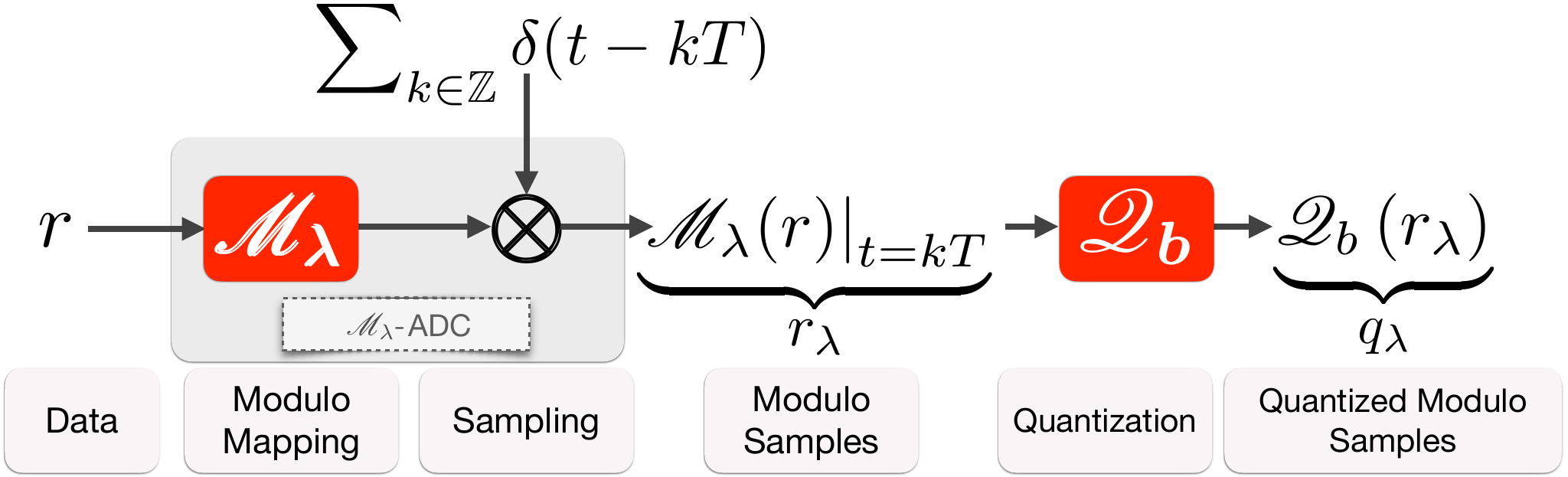}
		\caption{The architecture of the \madc and quantizer.}
		\label{fig:modarch}
	\end{center}
\end{figure}

Compared to the conventional setup, our newly designed receiver deploys a \madc to simultaneously reduce power consumption and prevents potential ADC saturation caused by limited dynamic range. As shown in \fig{fig:modarch},  an arbitrary baseband  signal $r(t)$ (continuous-time) is acquired by the \madc for both the $\mathsf{I}$/$\mathsf{Q}$ channels.  This is done by implementing the modulo mapping in hardware \cite{Bhandari:2021:J}. Formally, the centered modulo operator is defined by,
\begin{equation}
\mathscr{M}_{\lambda}: r \mapsto 2 \lambda\left(\biggr \llbracket \frac{r}{2 \lambda}+\frac{1}{2} \biggr \rrbracket-\frac{1}{2}\right),
\label{eq:modulo}
\end{equation}
where $\lambda>0$ is the folding  threshold and $\llbracket r \rrbracket \stackrel{\text { def }}{=} r-\lfloor r\rfloor$ is the fractional part of  $r$. 
If $|r(t)| < \lambda$, the \madc works the same as the conventional ADC.

Clearly, the action of \madc on $r(t)$ results in folding $r\rob{t}$ into low dynamic range signal that is upper bounded (in absolute sense), by $\lambda$; this also avoids the need for any VGA or AGC in the pipeline, thus facilitating lower complexity hardware. Finally, the continuous-time modulo-folded signal is sampled using impulse-train $$\otimes_{k T} \stackrel{\text { def }}{=} \sum\nolimits_{k \in \mathbb{Z}} \delta(t-k T), \quad T>0$$ 
where $T$ is the sampling interval. This results in modulo samples $\ym{r}{k}$,
\begin{equation}
\ym{r}{k} \stackrel{\text { def }}{=} r_\lambda(kT) =\mathscr{M}_\lambda(r(kT)), \quad k \in \mathbb{Z}.
\label{eq:mody}
\end{equation}

\bpara{Discrete-Time to Digital Modulo Samples.} Accounting for finite precision operations in practice, the modulo samples in \eqref{eq:mody} are quantized. Given a bit budget of $b$ bits, each modulo sample is rounded to the closest element in the set 
$$
\Upsilon_{b,\lambda} = \left\{\pm \frac{(2 n+1)}{2^b} \lambda \mid n \in\left\{0, \ldots, 2^{b-1}-1\right\}\right\}.
$$
Hence, the ``digital version'' of modulo samples is written as
\begin{equation}
    \label{eq:qb}
    \ym{q}{k} = \mathscr{Q}_b(\ym{r}{k}), 
    \quad
    \ym{q}{k} \in \Upsilon_{b,\lambda},
\end{equation}
where $\mathscr{Q}_b(\cdot)$ is the quantization mapping.
Specifically, the number of the mid-rise quantization levels is $B=2^b$ with the step size 
$\qo = 2^{-b} \rho_\lambda$ where $\rho_\lambda = 2\lambda$ is the dynamic range of the \madc. 

\begin{figure}[tb]
	\begin{center}
		\includegraphics[width =0.65\textwidth]{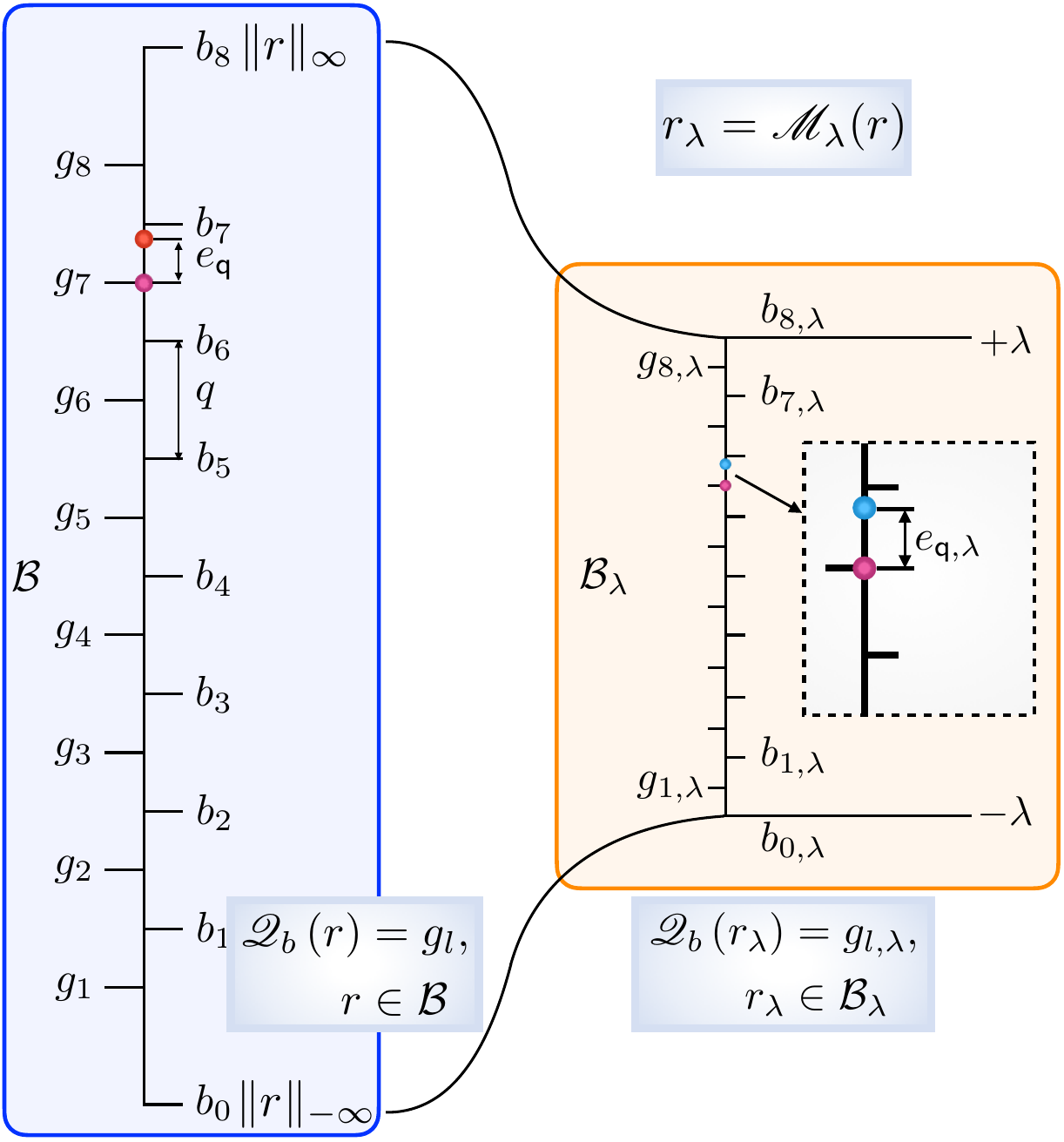}
		\caption{An illustration of quantization process of normal samples and modulo-folded samples with bit resolution $b=3$.}
		\label{fig:quant}
	\end{center}
\end{figure}

To measure the distortion caused by the quantization, we define the quantization noise as $e_\mathsf{q} \sqb{k} =  \ym{r}{k} - \mathscr{Q}_b(\ym{r}{k})$. As shown in \fig{fig:quant}, the synergistic interplay between the bit budget $b$ and dynamic range $\rho_\lambda$ decides the quantization noise $e_\mathsf{q}$. Given the same bit budget, it is clear that the quantization of modulo samples has \emph{smaller} quantization noise due to smaller dynamic range, determined by $\lambda$. The implication is that when compared to a conventional ADC, the \madc results in higher resolution of quantized samples. This advantage of the \madc over quantization on the normal samples \ie $r\rob{kT}$ is leveraged in the subsequent reconstruction.

\bpara{Recovery from Modulo Samples: The Unlimited Sampling Algorithm.} Reconstruction of the HDR input signal, $r\rob{t}$, from its modulo samples entails the ``inversion'' of the modulo non-linearity $\mathscr{M}_\lambda{\rob{\cdot}}$. This is an ill-posed problem due to the many-to-one nature of the $\mathscr{M}_\lambda{\rob{\cdot}}$. The key to unfolding modulo samples is the observation that the modulo and the finite-difference operators commute in a certain sense \cite{Bhandari:2020:Ja}. This allows us to extract the higher-order differences of $r$ from its modulo samples. Then, reconstruction boils down to the problem of stably inverting the difference-operator. Next, we focus on these two aspects separately and for the case of non-quantized measurements \eqref{eq:mody}. This is solely for the ease of exposition as the same recovery procedure applies to the case of quantized modulo samples in \eqref{eq:qb}.

\noindent \emph{Accessing Higher-Order Differences.} Let $\left( {\Delta r} \right)\left[ k \right] \DE r\left[ {k + 1} \right] - r\left[ k \right]$ denote the first-order difference operator; higher-order differences recursively defined by ${\Delta ^L}r \DE {\Delta ^{L - 1}}\left( {\Delta r} \right), L>1$. According to the modulo decomposition property (see Fig.~4 in \cite{Bhandari:2020:Ja}), we have, $r\rob{t} = \MO{r\rob{t}} + \VO{r}\rob{t}$ where $\VO{r}\rob{t}\in2\lambda\mathbb{Z}$ is a piecewise-constant function. The same property also applies to the point-wise samples. On the one hand,  since  $\VO{r}\rob{kT}\in2\lambda\mathbb{Z}$, we have that $\Delta^L \VO{r}\rob{kT}\in2\lambda\mathbb{Z}$ and hence, $\MO{\Delta^L \VO{r}\rob{kT}} = 0$. On the other hand, for $\Omega$-bandlimited functions, \ie $r\in\BL{}$, we note that $\Delta^L$ results in amplitude shrinkage in the oversampled regime. This is made precise via the following Lemma.  
\begin{lemma}[Difference-Derivative Inequality \cite{Bhandari:2017:C,Bhandari:2020:Ja}]
\label{Lemma}
For any $r \rob{t} \in \BL{} \cap {L}_\infty(\mathbb{{C}}) $, its samples $r\sqb{n}\DE r\rob{nT}$ satisfy, $\normt{\Delta^L r}{\infty}{\mathbb{R}} \leqslant \l T \Omega e 
\r ^L\normT{{r} }{\infty}{\mathbb{R}}$ where $e$ is Euler's number.
\end{lemma}
Clearly, choosing $T<1/\rob{\Omega e}$ amounts to controlling the factor $ \l T \Omega e 
\r ^L$. More precisely, choosing ${L^ \star } \geqslant \left\lceil {\frac{{\log \lambda  - {\beta _r}}}{{\log \left( {T\Omega e} \right)}}} \right\rceil$ with $\beta_r \geq 2\lambda\mathbb{Z}$ and $\beta_r \geq \normT{r}{\infty}{\mathbb{R}}$ also ensures that $\normt{\Delta^{L^\star} r}{\infty}{\mathbb{R}} \leqslant \lambda$.
According to Proposition 2 in \cite{Bhandari:2020:Ja}, we have, $\MO{\Delta^L\rob{r} } = \MO{\Delta^L \rob{\MO{r}} }$ and hence, whenever $L\geq L^{\star}$, oversampling or choosing $T\Omega e<1$ implies that $\Delta^L r = \MO{\Delta^L r_\lambda}$. In summary, oversampling $r\rob{t}\in \BL{}$ allows us to access $\Delta^L r$ from its modulo samples $r_\lambda\sqb{k}$. 

\noindent \emph{Stable Inversion of $\Delta^L$.} Instead of inverting $\rob{\Delta^L r}\sqb{k} \in \mathbb{R}$, the recovery algorithm in \cite{Bhandari:2017:C,Bhandari:2020:Ja} capitalizes the piece-wise constant structure of the residue $\VO{r}\rob{t}$, that is, $\VO{r}\rob{kT}\in2\lambda\mathbb{Z}$ and hence, $\Delta^L\VO{r}\rob{kT}\in2\lambda\mathbb{Z}$. This narrows down the range of $\VO{r}\rob{kT}$ and hence has a stabilizing effect during inversion. We initialize the sequence ${s^{\left\{ 0 \right\}}}\left[ k \right] = \rob{\MO{\Delta^L r_\lambda}-\Delta^L r_\lambda}\sqb{k} \equiv \Delta^L\VO{r}\rob{kT}\in2\lambda\mathbb{Z}$. Next, $\Delta^{L-1}\VO{r}\rob{kT}$ is obtained from $\Delta^L\VO{r}\rob{kT}$ by invoking the anti-difference operator $\mathsf{S}=\Delta^{-1}$ and rounded-off to $2\lambda\mathbb{Z}$. Note that the unknown constant-of-integration in the kernel of $\Delta$ must be estimated for unique reconstruction. It has been shown in \cite{Bhandari:2017:C,Bhandari:2020:Ja} that for $r\in\BL{}$, this unknown constant can be estimated via, 
\[
{\kappa ^{\left\{ l \right\}}} = \left\lfloor {\frac{{{\mathsf{S}}\left( {{s^{\left\{ {l + 1} \right\}}}} \right)\left[ k \right] - {\mathsf{S}}\left( {{s^{\left\{ {l + 1} \right\}}}} \right)\left[ {J_\lambda + 1} \right]}}{{12{\beta _r}}} + \frac{1}{2}} \right\rfloor ,\quad J_\lambda = 6\frac{{{\beta _r}}}{\lambda }, \quad l = 0,\ldots, L-2.
\]
This completes the estimation of ${s^{\left\{ 1 \right\}}}\left[ k \right]$ and more generally,  
\[{s^{\left\{ {l + 1} \right\}}}\left[ k \right] = 2\lambda \left\lceil {\frac{1}{2}\left\lfloor {\frac{{{\mathsf{S}}\left( {{s^{\left\{ {l + 1} \right\}}}} \right)\left[ k \right]}}{\lambda }} \right\rfloor } \right\rceil  + 2\lambda {\kappa ^{\left\{ l \right\}}}, \quad l = 0,\ldots, L-2. \]
Finally, we obtain the recovered samples $\widetilde{r}[k]$ up to an unknown constant,
\begin{equation}
    \widetilde{r}[k] = \mathsf{S} \left( s^{\{L-1\}}\right)[k]+\ym{r}{k}+2 \lambda \mathbb{Z}  =  \widetilde{\varepsilon}_r + \ym{r}{k}.
\end{equation}
The reconstruction approach is summarized in Algorithm~\ref{alg:modulorecovery} which also works with quantized measurements \cite{Bhandari:2020:Ja}. The recovery guarantee is summarized below.

\begin{algorithm}[t]
\label{alg:recovery}
\caption{Recovery from Noisy Modulo Communication Samples}
\label{alg:modulorecovery}
\textbf{Input:}  $\ym{q}{k}$ \text {and} $2 \lambda \mathbb{Z} \ni \beta_{r} \geqslant\|{r}\|_{\infty}$ . \\
\textbf{Output:}  $\widetilde{r}[k] \approx r(kT)$.
\begin{algorithmic}

\State \text {1) Compute} $L=\left\lceil\frac{\log \lambda-\log \beta_{r}}{\log (\mathrm{T} \Omega e)}\right\rceil$.
\State \text {2) Compute} $\left(\Delta^{L} q_\lambda\right)[k]$.
\State \text {3) Compute} $\left(\Delta^{L}\varepsilon_{r}\right)[k]=\left(\mathscr{M}_{\lambda}\left(\Delta^{L} q_\lambda\right)-\Delta^{L} q_\lambda\right)[k]$. 
\State \text {Set} $s^{\left\{ 0 \right\}}[k]=\left(\Delta^{L}\varepsilon_{r}\right)[k] $
\For{$l=0 : L-2$}
	\State \text {i)} $s^{\left\{ l+1 \right\}}[k]=\left(\mathsf{S} s^{\left\{ l \right\}}\right)[k]$
	\State \text {ii)} $s^{\left\{ l+1 \right\}}= 2 \lambda\left\lceil\frac{\left\lfloor s^{\left\{ l+1 \right\}} / \lambda \right \rfloor}{2}\right\rceil \text { (rounding to } 2 \lambda \mathbb{Z})$
	\State \text {iii) Compute} $ {\kappa ^{\left\{ l \right\}}}=\left\lfloor\frac{\left(\mathsf{S}^{2} \Delta^{l}\varepsilon_{r}\right)[1]-\left(\mathsf{S}^{2} \Delta^{l}\varepsilon_{r}\right)[J_\lambda+1]}{12 \beta_{r}}+\frac{1}{2}\right\rfloor$
	\State \text {iv)} $s^{\left\{ l+1 \right\}}[k]=s^{\left\{ l+1 \right\}}[k]+2 \lambda {\kappa ^{\left\{ l \right\}}}$
\EndFor
\State \text {5)} $\widetilde{r}[k]=\mathsf{S} \left(s_{\left\{ L-1 \right\}}\right)[k]+\ym{q}{k}+2 v \lambda, v \in \mathbb{Z}$
\end{algorithmic}
\end{algorithm}

\begin{theorem}[Unlimited Sampling Theorem with Bounded Noise \cite{Bhandari:2020:Ja}]
\label{thm:USTN}
Let $r(t)$ be a finite energy, bandlimited signal with maximum frequency $\Omega$ (rads/s) and assume that $\beta_r \in 2 \lambda \mathbb{Z}$ is known with $\|{r}\|_{\infty} \leqslant \beta_{r}$. For the dynamic range we work with the normalization $\rho_\lambda = \beta_{r}/\lambda$. Let the noise $\epsilon_\lambda$ be the additive form on the modulo samples with a noise bound given in terms of the dynamic range as 
\begin{equation}
\|\epsilon_\lambda\|_{\infty} \leqslant \frac{\lambda}{4}(2 \rho_\lambda)^{-\frac{1}{\alpha}}, \quad \alpha \in \mathbb{N}.
\end{equation}
Then a sufficient condition for Algorithm \ref{alg:modulorecovery} with
\begin{equation}
L =-\left\lceil\frac{\log \lambda-\log \beta_r-1}{\log (\mathrm{T} \Omega e)}\right\rceil,
\end{equation}
to approximately recover the bandlimited samples $r[k]$ in the sense that it returns $\widetilde{r}[k] = r[k] + \epsilon_\lambda + 2v\lambda$, $v \in \mathbb{Z}$ is that
\begin{equation}
\mathrm{T} \leqslant \frac{1}{2^\alpha \Omega e}.
\end{equation}
\end{theorem}

\section{Achievable sum-rate analysis}
\label{sec:asr}
To show the potential impact of the lower quantization noise achieved by \madc on the performance of communication, we analyze the SQNR and achievable sum-rate for both MRC and ZF combiners. Our analysis shows that with bits as few as $1$- or $2$-bits, the \lmimo architecture achieves very similar sum-rate performance when compared with $\infty$-bit resolution, conventional ADC.

To be able to compare with the benchmark (\ie conventional ADC), and clearly show the gain of \madc on the quality of quantization, we assume that the dynamic range $\rho_\lambda$ of the conventional ADC matches the signal $r$. Hence, the clipping error does not exist. Additionally, the quality of the quantization is measured by the signal to quantization noise ratio (SQNR) defined by \cite{gersho2012vector}
\begin{equation}
\mathsf{SQNR}=10 \log _{10} \frac{\sigma_{r}^{2}}{\sigma_{\mathsf{q}}^{2}},
\label{eq:sqnr}
\end{equation}
where $\sigma_{r}^{2}$ is the signal power with the assumption that signal $r$ has zero mean and $\sigma_{\mathsf{q}}^{2}$ is the quantization error power. 

As shown in \fig{fig:quant}, to calculate the $\sigma_{\mathsf{q}}^{2}$ at a given time instant during quantization, we first partition $\rho_\lambda$ into $J=B+1$ subsets expressed as $\rho_{\lambda,j}$, $j \in \iset{J}$. Each partition region is then bounded by the boundary values $b_j$, \ie $\rho_{\lambda,j}=\left[b_{j-1}, b_{j}\right)$. Specifically, the samples inside the same region $\rho_{\lambda,j}$ are mapped to the same quantization value $g_l$ based on the mapping \eqref{eq:qb}. Therefore, the variance $\sigma_\mathsf{q}^2$ of the quantization error is calculated as \cite{gersho2012vector}, \cite{wang2002video}
\begin{equation}
\begin{aligned}
\sigma_\mathsf{q}^{2}&= \mathscr{E}\left\{|r-\mathscr{Q}_b(r)|^{2}\right\} 
=  \sum\limits_{j \in \iset{J}}  P\left(r \in \rho_{\lambda,j}\right)\int_{b_{j-1}}^{b_{j}}  \left(r-g_{j}\right)^{2}  p\left(r \mid r \in \rho_{\lambda,j}\right) d y,
\label{eq:sigmaq}
\end{aligned}
\end{equation}
where 
\begin{enumerate}[leftmargin = *, label = ---]
    \item $p(r)$ is probability density function (pdf) of $r$.
    \item $P\left(r \in \rho_{\lambda,j}\right)=\int_{\rho_{\lambda,j}} p(r) d r$ is the probability when signal $r$ is inside the $\rho_{\lambda,j}$ region \ie{$r \in \rho_{\lambda,j}$}.
    \item $p\left(r \mid r \in \rho_{\lambda,j}\right)=p(r)/P(r \in \rho_{\lambda,j})$ is the conditional pdf of $r$ in region $\rho_{\lambda,j}$.
\end{enumerate}

 In comparison, the quantization error of \madc is decreased to $\sigma_{\lambda,\mathsf{q}}^2=\zeta^2\sigma_\mathsf{q}^2$ due to modulo operation. The ratio $\zeta$ is defined by
 \begin{equation}
     \zeta = \frac{\lambda}{\| r \|_\infty},
     \quad \| r \|_\infty = \max |r\rob{t}|.
 \end{equation}
With the above, we derive the SQNRs for baseband sampled single carrier and OFDM signals as follows.

\bpara{SQNR of Uniform Distribution Signal.} The baseband sampled single-carrier signal is subject to uniform distribution that $r[k] \sim  \mathcal{U}(0,1)$, hence the variance of the sample $r[k]$ is $\sigma_{r,\mathsf{unif}}^{2} = \rho_\lambda^2/12$, and variance of quantization noise $\sigma_\mathsf{q, unif}^{2}$ is given by
\begin{equation}
\sigma_\mathsf{q, unif}^{2}=\frac{q_0^{2}}{12}=\sigma_{r,\mathsf{unif}}^{2} \ 2^{-2 b}.
\end{equation}
Subsequently, the SQNR of conventional ADC is 
\begin{equation}
\mathsf{SQNR}_\mathsf{unif}=10 \log _{10} \frac{\sigma_{r,\mathsf{unif}}^{2}}{\sigma_{\mathsf{q,\mathsf{unif}}}^{2}}=6.02 b \ (\mathrm{dB}).
\label{eq:sqnr1c}
\end{equation}
Since the modulo-folded samples follow uniform distribution that $r_\lambda[k] \sim \mathcal{U}(0, \zeta^2)$, the quantization error of \madc is decreased to $\sigma_{\mathsf{q, \lambda}}^2=\zeta^2\sigma_\mathsf{q,unif}^2$ with the SQNR given by
\begin{equation}
\mathsf{SQNR}_\mathsf{unif, \lambda}=10 \log _{10} \frac{\sigma_{r, \mathsf{unif}}^{2}}{\sigma_{\mathsf{q,\lambda}}^{2}}=6.02 b + 2 \log _{10} (1/\zeta)  \ (\mathrm{dB}).
\label{eq:sqnr1}
\end{equation}
Therefore, there is a $2 \log _{10} (1/\zeta)$ dB enhancement on SQNR by adopting \madc.

\begin{figure}[tb]
\centering
		\includegraphics[width =0.75\textwidth]{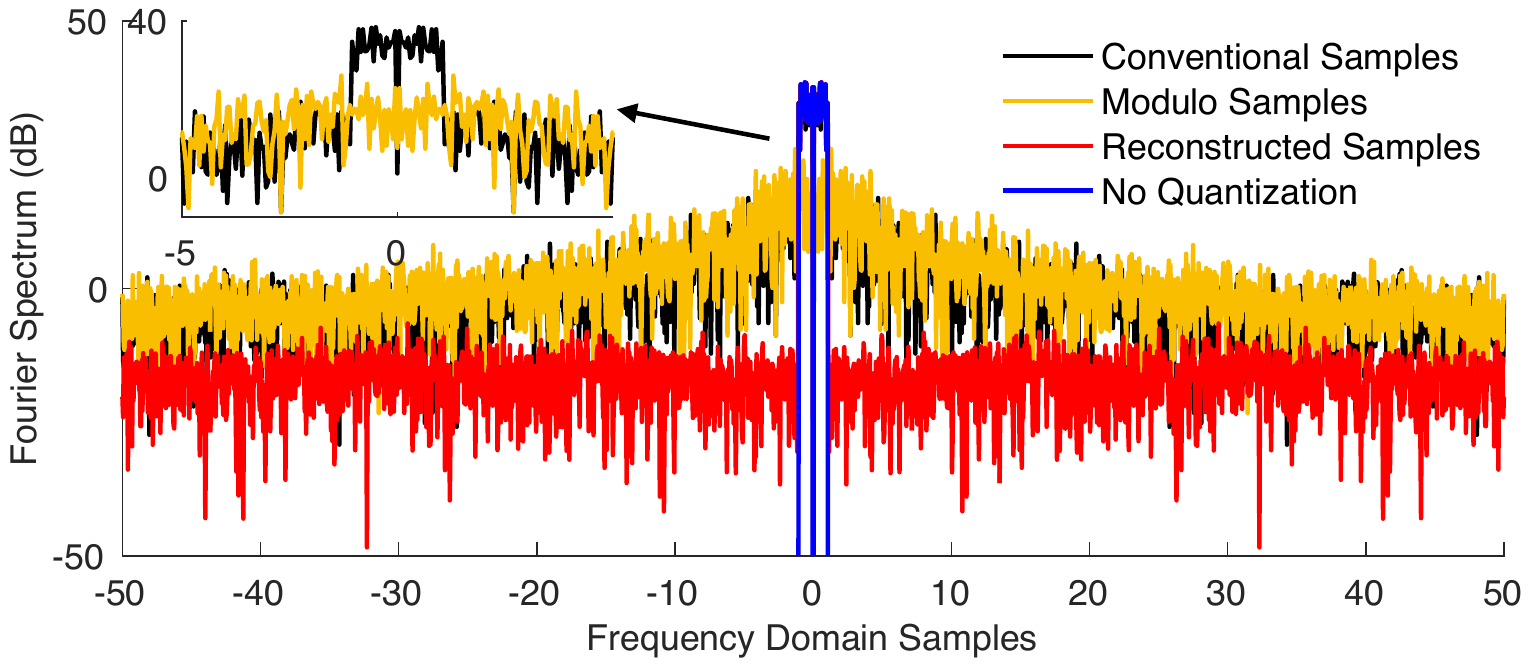}
		\caption{Fourier spectrum of $r$, and 4-bit quantized samples for conventional ADC and \madc with $\zeta = 0.1$.}
	    \label{fig:quant_noise}
\end{figure}

\bpara{SQNR of Gaussian Distribution Signal.}\label{subsec:metric} The OFDM signal is commonly assumed to be a centered, wide-sense stationary, ergodic Gaussian process \cite{dardari2006joint}, \cite{bernhard2012analytical} that is $r[k] \sim \mathcal{N}(0,1)$, thus the variance of quantization noise $\sigma_\mathsf{q, Gau}^{2}$ is given by
\cite{gersho2012vector,wang2002video, panter1951quantization}
\begin{equation}
\sigma_\mathsf{q, Gau}^{2}=\varepsilon^{2} \sigma_{r, \mathsf{Gau}}^{2} 2^{-2 b},
\end{equation}
where
\begin{equation}
\varepsilon^{2}=\frac{1}{12 \sigma_{r, \mathsf{Gau}}^{2}}\left(\int_{-\infty}^{\infty} \sqrt[3]{p(r)}d r\right)^{3}.
\end{equation}
Specifically, for the case when $r[k] \sim \mathcal{N}(0,1)$, we have $\varepsilon^{2} =\sqrt{3}\pi / {2}$.
Thus, the SQNR of the conventional ADC is given by
\begin{equation}
    \mathsf{SQNR}_\mathsf{Gau} = 10 \log _{10} \rob{\frac{\sigma_{r, \mathsf{Gau}}^{2}}{\sigma_{\mathsf{q, Gau, (m)}}^{2}}}
     \approx  6.02 b  - 4.35 \ (\mathrm{dB}).
    \label{eq:sqnrgauc}
\end{equation}
Modulo folding on the input distribution leads to a family of so-called ``wrapped distributions'' \cite{mardia2000directional}. In particular, when the input signal is drawn from $\mathcal{N}\rob{0,\sigma^2}$ and the modulo threshold $\lambda$ is chosen to be a fraction of $\sigma^2$, empirically, we have found that the resulting density function for $r_\lambda[k]$ follows a uniform distribution. Assuming that $r_\lambda[k] \sim \mathcal{U}(0, \zeta^2)$, the quantization error of \madc shrinks to $\sigma_{\mathsf{q, \lambda}}^{2}\approx\zeta^2\sigma_\mathsf{q,unif}^2$, and the SQNR is given by 
\begin{equation}
    \begin{aligned}
    \mathsf{SQNR}_\mathsf{Gau, \lambda}& = 10 \log _{10} \rob{\frac{\sigma_{r, \mathsf{Gau}}^{2}}{\sigma_{\mathsf{q, \lambda}}^{2} }} \approx  6.02 b  + 2 \log _{10} (1/\zeta)  \ (\mathrm{dB}).
    \end{aligned}
    \label{eq:sqnrgaum}
\end{equation}
In summary, when capitalizing on the \madc the resulting samples entail lower quantization noise as shown in \fig{fig:quant_noise}. This in turn leads to a factor $2 \log _{10} (1/\zeta) + 10 \log_{10}(\varepsilon^{2})$ dB enhancement in the SQNR. For bandlimited functions $r(t)$ with maximum frequency $\Omega$, from \cite{Bhandari:2017:C,Bhandari:2020:Ja} we know that ${\left\| {{\Delta ^L}r} \right\|_\infty } \leqslant {\left( {T\Omega  e} \right)^L}{\left\| r \right\|_{{L_\infty }}}$. Choosing $T<1/\rob{\Omega e}$ for an appropriate value of $L$ entails that,
\[{\left( {T\Omega e} \right)^L} < \frac{\lambda }{{{{\left\| r \right\|}_{{L_\infty }}}}} = \zeta.\]
For example, choosing $\zeta \in \mathcal{U} \left({1}/{100}, {1}/{10}\right)$ we can see that SQNR \eqref{eq:sqnrgaum} can be increased from $24.5$ dB to $44.5$ dB (in noiseless conditions). 

\bpara{Achievable sum-rate Analysis for \madc in MU \lmimo.}\label{sec:sr} To be able to show the benefits of using \madc in communication systems, we analyze the achievable sum-rate of MU \lmimo architecture. The non-linear nature of the quantization operator makes the analysis challenging. To overcome this aspect, we adopt the AQNM model, previously described in \cite{mo2015capacity, fan2015uplink, orhan2015low}. Consequently, the quantized signal $\mathbf{r}_\mathsf{q}$ is modeled by, 
\begin{equation}
\mathbf{r}_{\mathsf{q}}=\gamma_\lambda \mathbf{r}+\boldsymbol{\epsilon}_{\mathsf{q}}=\gamma_\lambda \sqrt{p_\mathrm{u}} \mathbf{H} \mathbf{x}+\gamma_\lambda \boldsymbol{\epsilon}+\boldsymbol{\epsilon}_{\mathsf{q}},
\label{eq:aqnm}
\end{equation}
where $p_\mathrm{u}$ is the average transmitted power of each user, $\gamma_\lambda = 1 - {\sigma_\mathsf{q,\lambda}^{2}}/{\sigma_{r}^{2}}$, $\boldsymbol{\epsilon}_{\mathsf{q}} \sim \mathcal{N}(0, (1-\gamma_\lambda)\gamma_\lambda\sigma_{r}^{2})$ is the additive Gaussian quantization noise independent with $\mathbf{r}$. In this setting, the signal vector after the linear detection is given by
\begin{equation}
    \widetilde{\mathbf{x}}_{\mathsf{b}} = \mathbf{A}^H \mathbf{r}_\mathsf{q},
\label{eq:ay}
\end{equation}
where $\mathbf{A}$ is given in \eqref{eq:combiner}. By substituting \eqref{eq:ay} into \eqref{eq:aqnm}, we derive the following expression
\begin{equation}
\widetilde{\mathbf{x}}_{\mathsf{b}} =\gamma_\lambda \sqrt{p_\mathrm{u}} \mathbf{A}^H \mathbf{H} \mathbf{x}+\gamma_\lambda \mathbf{A}^H \boldsymbol{\epsilon}+\mathbf{A}^H \boldsymbol{\epsilon}_{\mathsf{q}}.
\label{eq:aqnm_v}
\end{equation}
Then, the $m^\textrm{th}$ user data is $[\widetilde{\mathbf{x}}_{\mathsf{b}}]_m = \widetilde{{x}}_{m, \mathsf{b}}$ is given by
\begin{equation}
\begin{aligned}
\widetilde{{x}}_{m, \mathsf{b}}&=\sqrt{p_\mathrm{u}} \gamma_\lambda x_{m,\mathsf{b}}
\left\langle {{{\mathbf{a}}_m},{{\mathbf{h}}_m}} \right\rangle +\sqrt{p_\mathrm{u}} \gamma_\lambda \sum \limits _{\substack{i=1 \\ i \neq m}}^{M} x_{i, \mathsf{b}} \left\langle {{{\mathbf{a}}_m},{{\mathbf{h}}_i}} \right\rangle 
+\gamma_\lambda \left\langle {{{\mathbf{a}}_m},\boldsymbol{\epsilon}} \right\rangle +  \left\langle {{{\mathbf{a}}_m},\boldsymbol{\epsilon}_{\mathbf{q}}} \right\rangle,
\end{aligned}
\label{eq:vn}
\end{equation}
where $\mathbf{a}_{m}$ is the $m^\textrm{th}$ column of $\mathbf{A}$. From \eqref{eq:vn}, we can also obtain the interference and noise term by excluding the $m^\textrm{th}$ user's date, given by
\begin{equation}
\begin{aligned}
\mathrm{I}_{\mathbf{H}}=& p_\mathrm{u} \gamma_\lambda^{2} \sum_{\substack{i=1 \\ i \neq m}}^{N}\left|
\left\langle {{{\mathbf{a}}_m},{{\mathbf{h}}_i}} \right\rangle \right|^{2}+\gamma_\lambda^{2}\left\|\mathbf{a}_m\right\|^{2} + 
\left\langle \mathbf{a}_m,\mathbf{R}_{\boldsymbol{\epsilon}_{\mathsf{q}},  
\boldsymbol{\epsilon}_{\mathsf{q}} }\mathbf{a}_{m} \right\rangle 
\end{aligned}
\end{equation}
where $\mathbf{R}_{\boldsymbol{\epsilon}_{\mathsf{q}},  \boldsymbol{\epsilon}_{\mathsf{q}}}$ is the covariance of the additive quantization noise $\boldsymbol{\epsilon}_{\mathsf{q}}$ in a given channel realization with
\begin{equation}
\mathbf{R}_{\boldsymbol{\epsilon}_{\mathsf{q}},  \boldsymbol{\epsilon}_{\mathsf{q}}}=\gamma_\lambda(1-\gamma_\lambda) \operatorname{diag}\left(p_\mathrm{u} \mathbf{H} \mathbf{R}_\mathbf{x} \mathbf{H}^{H}+\mathbf{I}\right).
\label{eq:rnq}
\end{equation}
Specifically, $\mathbf{R}_\mathbf{x}$ is the input signal covariance, and $\mathbf{R}_\mathbf{x} = \mathbf{I}$ when $r[k] \sim \mathcal{N}$. 
The above derivations allow us to obtain the ergodic achievable uplink rate of the $m^\textrm{th}$ user \cite{fan2015uplink} that is,
\begin{equation}
\label{eq:ergouprate}
R_m=\mathscr{E}\left\{\log _{2}\left(1+\frac{p_\mathrm{u} \gamma_\lambda^{2}\left\|\mathbf{a}_m\right\|^{4}}{\mathrm{I}_{\mathbf{H}}}\right)\right\}.
\end{equation}
Since the direct computation of the achievable uplink rate in \eqref{eq:ergouprate} is challenging, in the sequel, we will utilize an analytical approximation for the \lmimo case.

\subsubsection{MRC Analytical Approximation of Achievable sum-rate}

The uplink rate of the $m^\textrm{th}$ user with MRC is given by \cite{fan2015uplink}
\begin{equation}
\widetilde{R}_{m, \mathsf{(MRC)}}=\log _{2}\left(1+\frac{p_\mathrm{u} \gamma_\lambda \eta_m(N+1)}{\mathrm{I}_\mathsf{MRC}}\right),
\label{rateuser}
\end{equation}
where $\mathrm{I}_\mathsf{MRC}$ is the expectation of the noise-plus-interference term with
\begin{equation}
\mathrm{I}_\mathsf{MRC}=p_\mathrm{u} \gamma_\lambda \sum\limits_{\substack{i=1 \\i \neq n}}^{M} \eta_{i}+p_\mathrm{u}(1-\gamma_\lambda)\left(\sum_{i=1}^{M} \eta_{i}+\eta_m\right)+1,
\label{jrateuser}
\end{equation}
and the sum-rate for the system is $R=\sum_{m=1}^{M} \widetilde{R}_m$.

\subsubsection{ZF Analytical Approximation of Achievable sum-rate}

Similarly, the uplink rate of the $m^\textrm{th}$ user with ZF is given by \cite{qiao2016spectral}
\begin{equation}
\tilde{R}_{m, \mathsf{(ZF)}}=\log _{2}\left(1+\frac{\gamma_\lambda p_\mathrm{u}}{\frac{\gamma_\lambda}{(N-M) \eta_m}+(1-\gamma_\lambda) \mathrm{I}_\mathsf{ZF}}\right),
\end{equation}
where $\mathrm{I}_\mathsf{ZF}$ is the expectation of the noise-plus-interference term given in \eqref{eq:izf}, $\delta < 1$ is the attenuation coefficient.
\begin{figure*}
\begin{equation}\label{eq:izf}
\begin{aligned}
\mathrm{I}_\mathsf{ZF}& = N\left(\frac{\delta}{N+M}\right)^{2}\left\{4 \eta_{m}-\frac{4 \delta}{N+M} \eta_{m}\left(\sum_{j=1}^{M} \eta_{j}+\eta_{m}\right)+N \eta_{m}\left(\frac{\delta}{N+M}\right)^{2}\left(\sum_{j=1}^{M} \eta_{j}^{2}+\eta_{m}^{2}\right)\right.\\
&+p_\mathsf{u} \sum_{i=1, i \neq k}^{M}\left[4 \eta_{m} \eta_{i}-\eta_{m} \eta_{i} \frac{4 \delta}{N+M}\left(\sum_{j=1}^{M} \eta_{j}+\eta_{m}+\eta_{i}\right)+N \eta_{m} \eta_{i}\left(\frac{8}{N+M}\right)^{2}\left(\sum_{j=1}^{M} \eta_{j}^{2}+\eta_{m}^{2}+\eta_{i}^{2}\right)\right] \\
&\left.+8 p_\mathsf{u} \eta_{m}{ }^{2}-p_\mathsf{u} \eta_{m}{ }^{2} \frac{8 \delta}{N+M}\left(\sum_{j=1}^{M} \eta_{j}+2 \eta_{m}\right)+p_\mathsf{u} \eta_{m}^{2}\left(\frac{\delta}{N+M}\right)^{2}\left((N+1) \sum_{j=1}^{M} \eta_{j}^{2}+(3 N+1) \eta_{m}^{2}\right)\right\}.
\end{aligned}
\end{equation}
\hrulefill
\end{figure*}

\section{Numerical evaluation}
\label{sec:result}

The focus of this section is to demonstrate the advantages of our proposed \lmimo approach. Our experiments are based on both computer simulations as well as modulo ADC hardware. 

Starting with computer experiments, we show the recovery and detection performance of the high-order modulation schemes (\eg $1024$ QAM OFDM signal) when using the \madc. The SQNRs of the uniform and Gaussian distributed baseband samples in \madc and conventional ADC are compared with theory in Section \ref{subsec: recovery}. We show proof-of-concept hardware experiments in Section \ref{subsec:exp}. Additionally, the ergodic achievable sum-rate of \madc with MRC and ZF detectors, and the analytic approximations are discussed in Section \ref{subsec: asr}. 

\subsection{Recovery and Detection Performance}
\label{subsec: recovery}
We consider two cases that \nbsc and \mimoo in the simulation. Specifically, \nbsc is with channel tap $D=1$, small-scale fading coefficients $g_{{n,m}} \sim \mathcal{CN}(0, 1)$. The \mimoo considers the frequency selective channel with channel taps $D=15$ with small-scale fading coefficient\footnote{Uniform power delay profile is assumed for simplicity of illustration. More realistic channel models can be used} $g_{{n,m}}[d] \sim \mathcal{CN} (0, 1/D)$. We assume that perfect CSI, maximum value of baseband signal $\|r\|_{\infty}$, $\lambda=\frac{1}{10}\|r\|_{\infty}$ are known at the receiver. To satisfy the sufficient condition for recovery, we use $f_s \geq 2 \pi e f_\mathsf{Nyquist}$ in the simulation.

\begin{figure}[tb]
\centering
\includegraphics[width =0.65\textwidth]{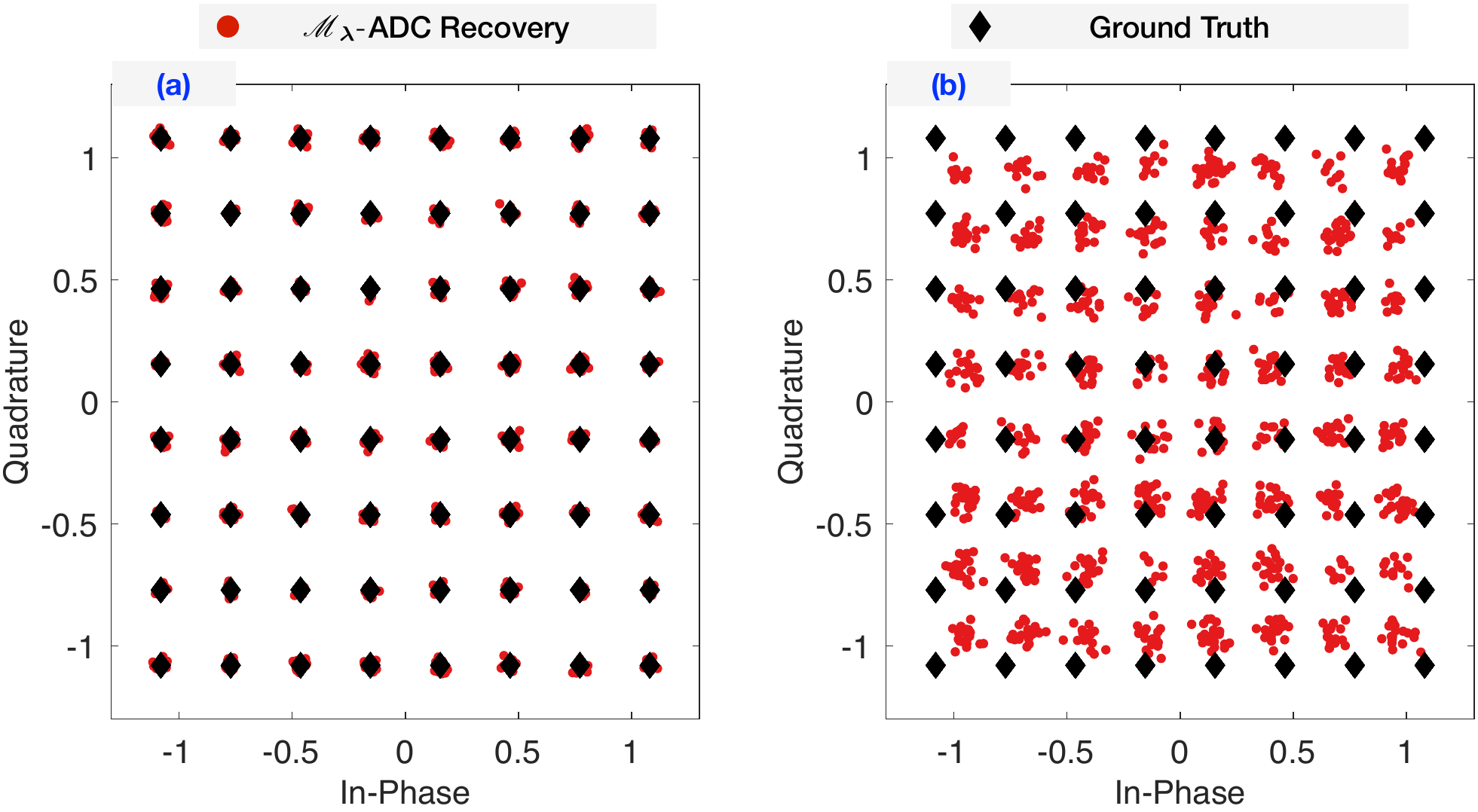}
\caption{Constellation diagrams. (a) $N=10$ users with $M=50$ BS antennas and ZF detector. (b) $N=5$ users with $M=500$ BS antennas and MRC detector.}
\label{fig:cons}
\end{figure}

\begin{figure*}[tb]
\centering
		\includegraphics[width =1\textwidth]{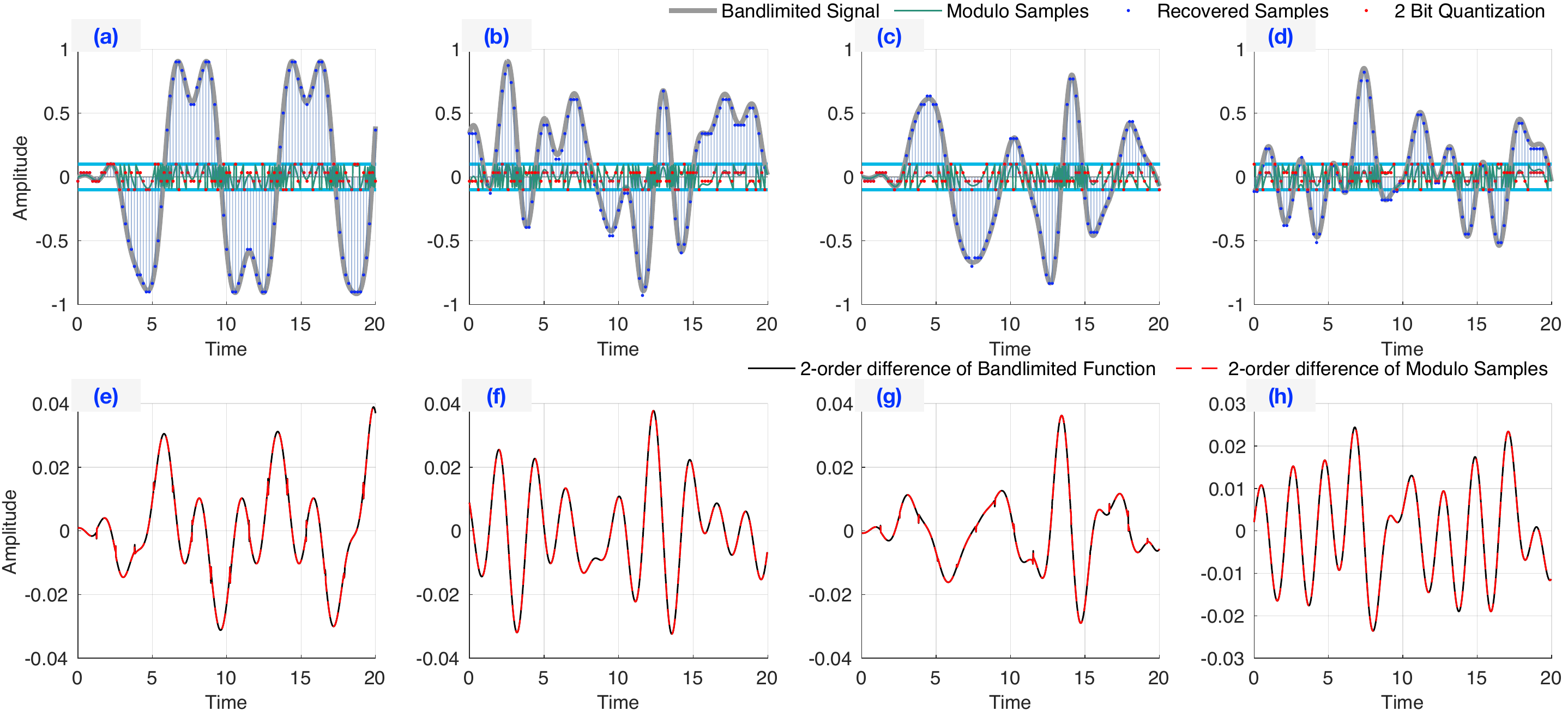}
		\caption{Modulo samples and recovery of QPSK (a) single-carrier signal, (b) OFDM signal, $1024$ QAM (c) single-carrier signal, (d) OFDM signal; 2-order difference on samples and modulo-folded samples of QPSK (e) single-carrier signal, (f) OFDM signal, $1024$ QAM (g) single-carrier signal, (h) OFDM signal.}
		\label{fig:sc_OFDM_recovery}
\end{figure*}

At the receiver with \madc, the received signals are modulo folded, sampled, quantized, recovered, and combined by ZF or MRC detectors. In \fig{fig:cons}, the ZF and MRC receiver with \madcs outputs are plotted for the case that single-antenna multiple users transmit $64$QAM modulation signals. The observations are that the recovered signals from modulo-folded samples can be detected by normal linear combiners, and a better mapping in the constellation can be achieved by increasing the antenna number in the BS.

In \fig{fig:sc_OFDM_recovery} (a), (b), the recovery results of one user's data for the cases that users transmit QPSK single-carrier and QPSK OFDM signals are demonstrated. In addition, in \fig{fig:sc_OFDM_recovery} (c), (d), the $1024$ QAM single-carrier and OFDM signals after the same processes are shown. The recovery performance is also analyzed by reconstruction mean squared error (MSE) between input signal $r$ and its estimate $\tilde{r}$ recovered by Algorithm \ref{alg:modulorecovery}, symbol error rate (SER), and bit error rate (BER). Specifically, the result of the recovery single-carrier signal with $1024$ QAM is that MSE $=3.8 \times 10^{-4}$, BER $=7.7 \times 10^{-4}$ and SER $=7.9 \times 10^{-3}$. Using the same simulation parameters, OFDM signal with 1024 QAM can also be fully recovered with MSE $= 6.4 \times 10^{-4}$, BER $=3.0 \times 10^{-3}$ and SER $=1.7 \times 10^{-2}$. 

From the \fig{fig:cons} and \fig{fig:sc_OFDM_recovery}, it is known that the once incapable recovery and detection of very high order modulation signal can be realized with only 2 bit resolution \madc with $f_\mathrm{s} = 50 f_\mathsf{Nyquist}$. According to the \fig{fig:sc_OFDM_recovery} (e), (f) and \fig{fig:sc_OFDM_recovery} (g), (h), we can know $\Delta^{L} r=\mathscr{M}_{\lambda}\left(\Delta^{L} q_\lambda\right)$, which means the modulo samples of higher order of bandlimited samples $r$ and modulo samples $q_\lambda$ are the same. This fact provides the evidence of the unlimited sampling theory's effectiveness on communication signals.

When the bit resolution of \madc increases from $b=2$ to $12$, the MSE of $1024$ QAM single-carrier signal decreases from $3.8 \times 10^{-4}$ to $2.0 \times 10^{-10}$. Additionally, the MSE of $1024$ QAM OFDM signal decreases from $6.4 \times 10^{-4}$ to $2.0 \times 10^{-10}$. These results show that the recovery results of $\widetilde{r}$ based on Algorithm \ref{alg:modulorecovery} become better. Considering a noise case, the detection performances of $1024$ QAM signal are BER $=9.0543 \times 10^{-4}$ and SER $=9.1 \times 10^{-3}$ with SNR $=55$ dB. The improvement of the noise resistance of the Algorithm \ref{alg:modulorecovery} is left in the future work.

\begin{figure}[tb]
\centering
		\includegraphics[width =0.65\textwidth]{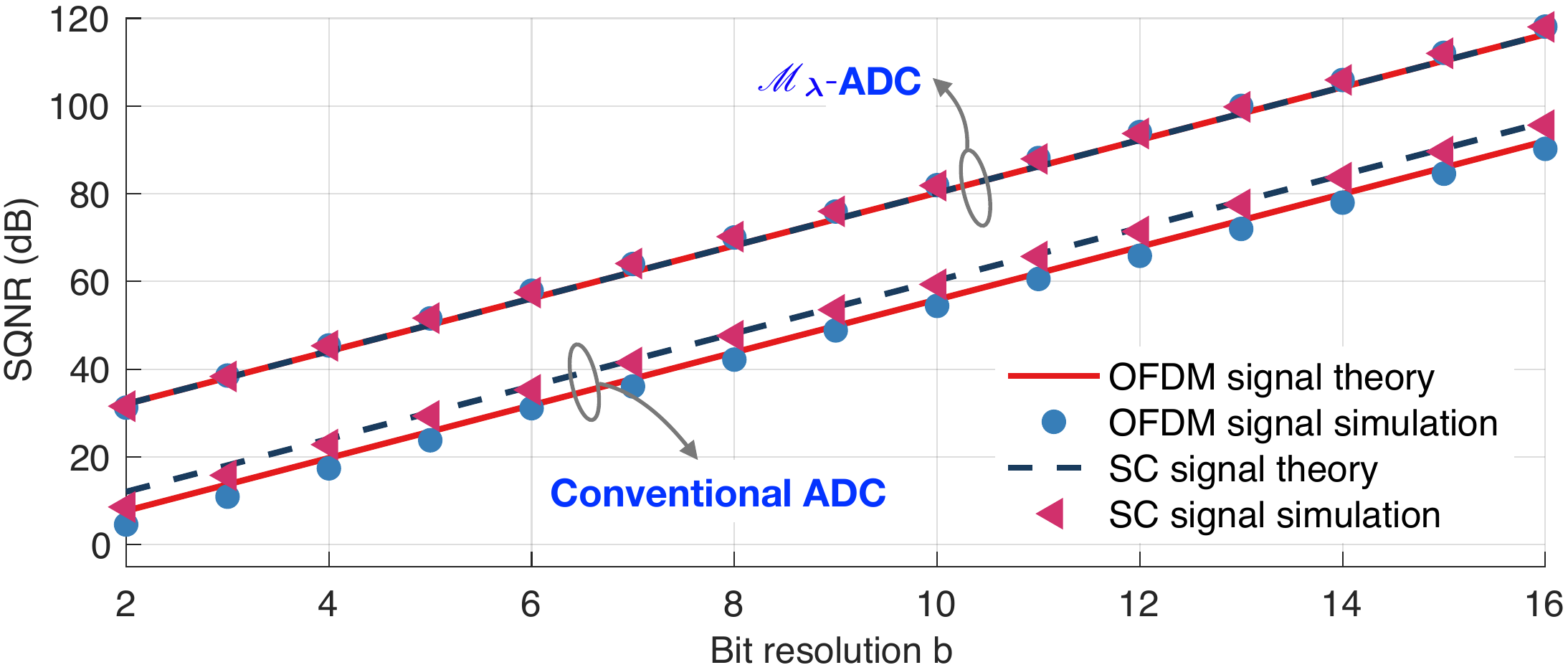}
		\caption{SQNR of $\mathscr{M}_{\lambda}$ and conventional ADCs with $\lambda = 0.1  \|r\|_{\infty}$. }
	    \label{fig:sqnr}
\end{figure}

\begin{figure}[tb]
\centering
\includegraphics[width =0.75\textwidth]{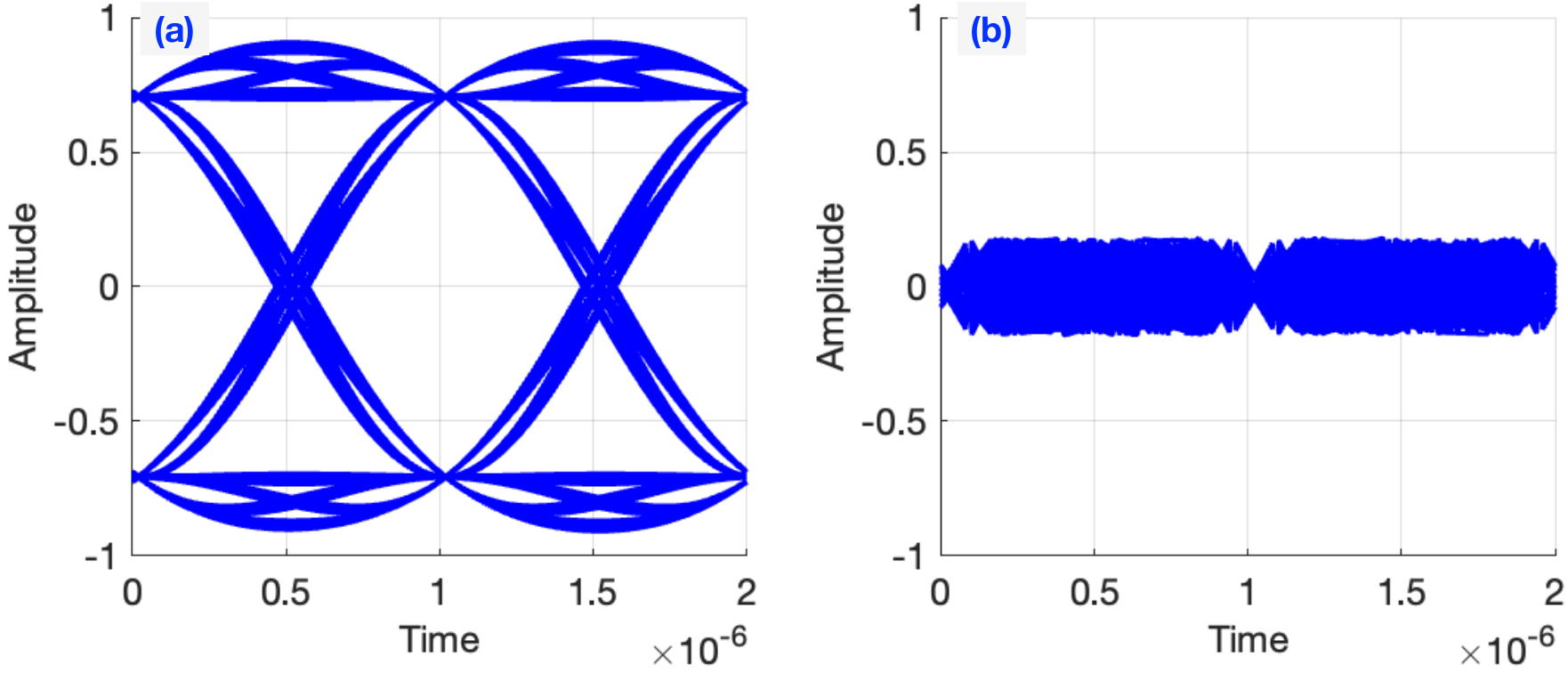}
\caption{Eye diagrams with roll-off factor $\alpha = 0.5$ for signals (a) after reconstruction, (b) before reconstruction (modulo signal).}
\label{fig:eye}
\end{figure}

\begin{figure*}[tb]
	\begin{center}
		\includegraphics[width =1\textwidth]{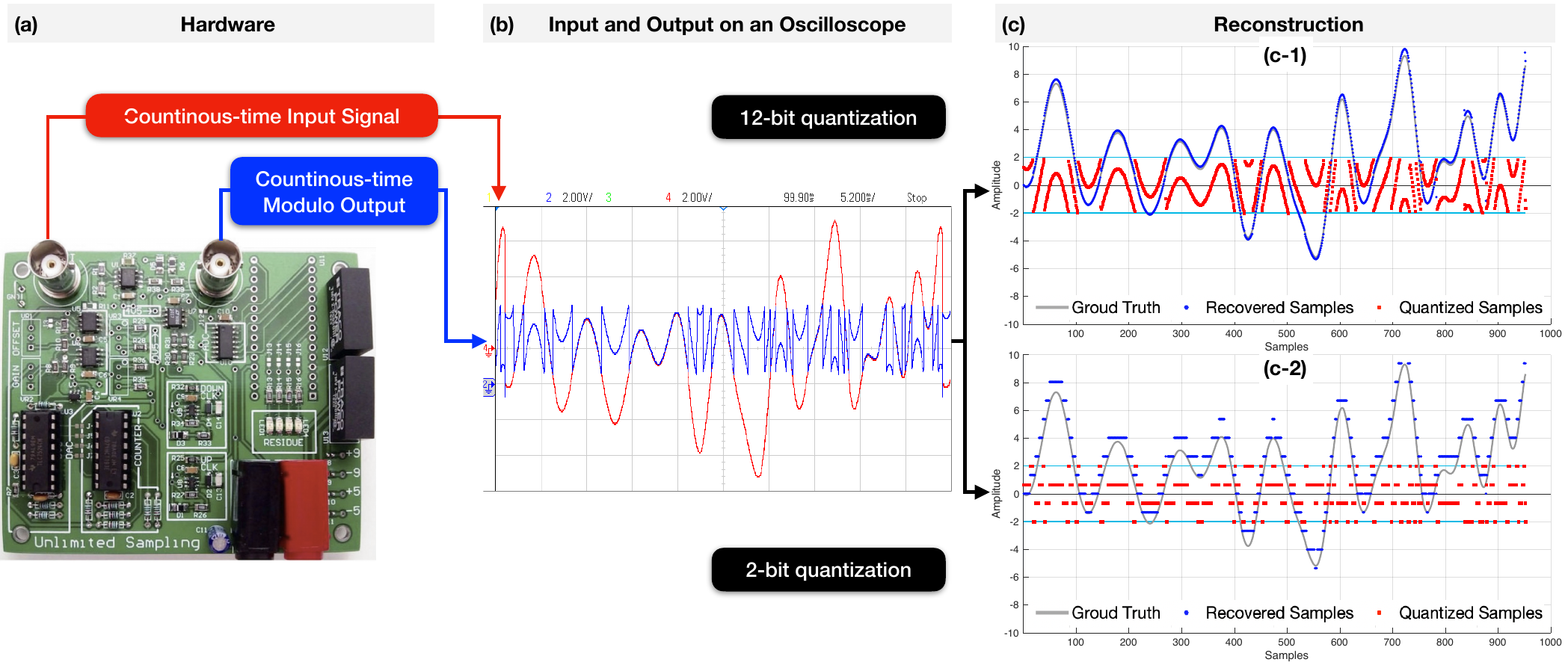}
		\caption{Hardware experiments and actual reconstruction results. (a) \madc circuit. (b) Live screen shot of the oscilloscope plotting the real part of QPSK signal (red) and its modulo-folded signal (blue). (c) Reconstruction results with actual modulo-folded signals from the circuit as the input, and 2-bit and 12-bit resolution quantization.}
		\label{fig:hd}
	\end{center}
\end{figure*}

In addition, the SQNRs of single-carrier and OFDM signals with different ADC resolutions $b$ are shown in \fig{fig:sqnr}. It is clear that a higher SQNR is achieved by using \madc than the conventional ADC. We also show the eye diagrams of signal after and before reconstruction in \fig{fig:eye}. The distortion of the reconstructed signal is low, which validates the effectiveness of the reconstruction algorithm.

\subsection{Hardware Experiment}
\label{subsec:exp}
We show proof-of-concept hardware experiments of \madc in this section. Here, we use modulo samples of the real component of the QPSK signal. Specifically, the settings are $f_s = 50 f_\mathsf{Nyquist}$, dynamic range $13.6$V peak-to-peak ($\approx 7 \lambda$), and folded dynamic range $4.02$V peak-to-peak. The experimental data and recovery results are shown in \fig{fig:hd}. We can know that the original real component of the QPSK signal can be recovered with $\textrm{MSE} =4.6 \times 10^{-2}$ and $7.5 \times 10^{-1}$ when use $12$-bit and $2$-bit quantization resolution, respectively.

\subsection{Achievable Uplink sum-rates and Analytical Approximations}
\label{subsec: asr}

\begin{figure}[tb]
	\begin{center}
		\includegraphics[width = 1\textwidth]{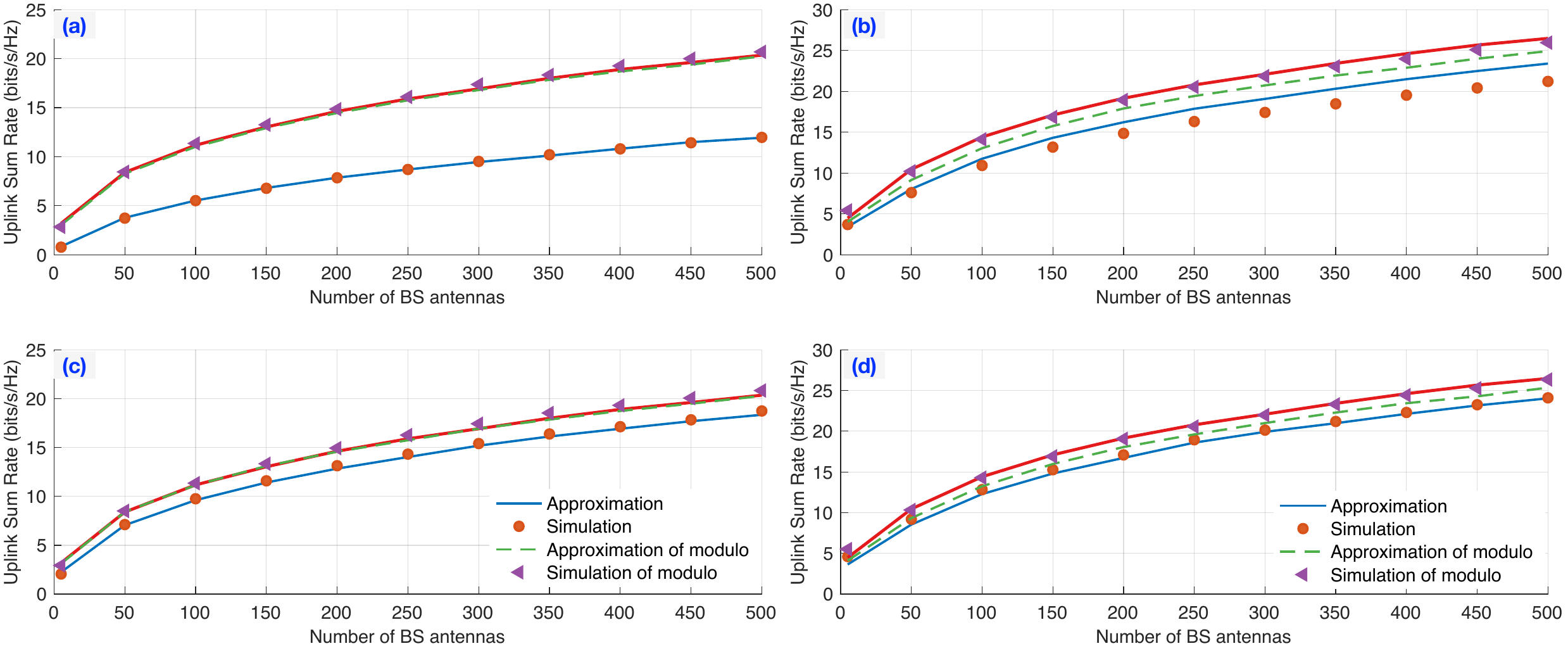}
		\caption{Uplink sum-rate per cell vs the number of BS antennas, when $N=10$ users with Tx power $p_\mathsf{u}=10$ dB and BS using (a), (c) MRC detector; (b), (d) ZF detector.}
		\label{fig:mrcAndzfsr}
	\end{center}
\end{figure}

\begin{figure}[tb]
	\begin{center}
		\includegraphics[width =0.65\textwidth]{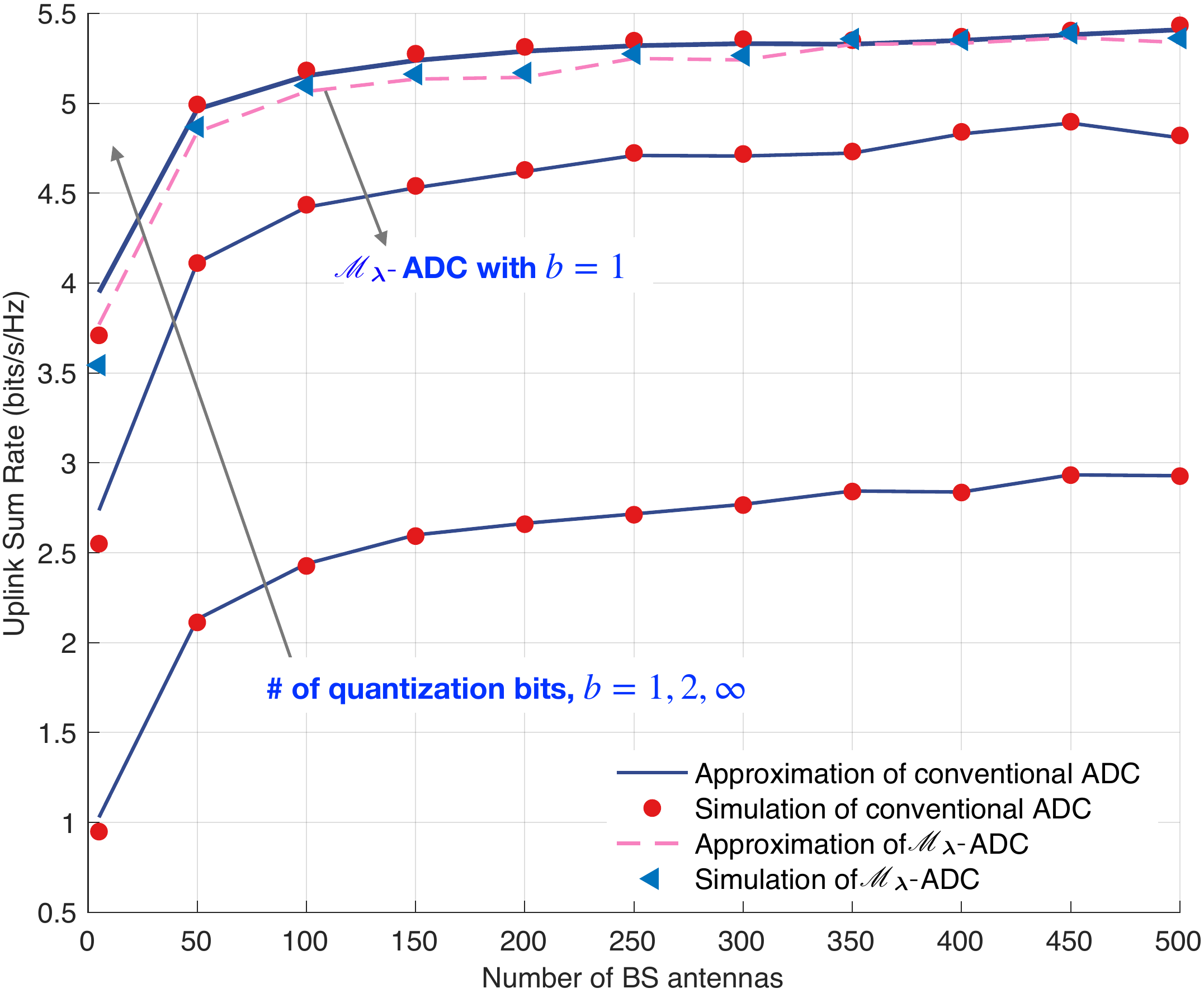}
		\caption{Uplink sum-rate per cell vs the number of BS antennas, when $N=10$ users with scaled-down Tx power $p_\mathsf{u}=\frac{E_\mathsf{u}}{M}$ dB and BS using MRC detector.}
		\label{fig:sreumrc}
	\end{center}
\end{figure}

\begin{figure}[tb]
	\begin{center}
		\includegraphics[width =0.65\textwidth]{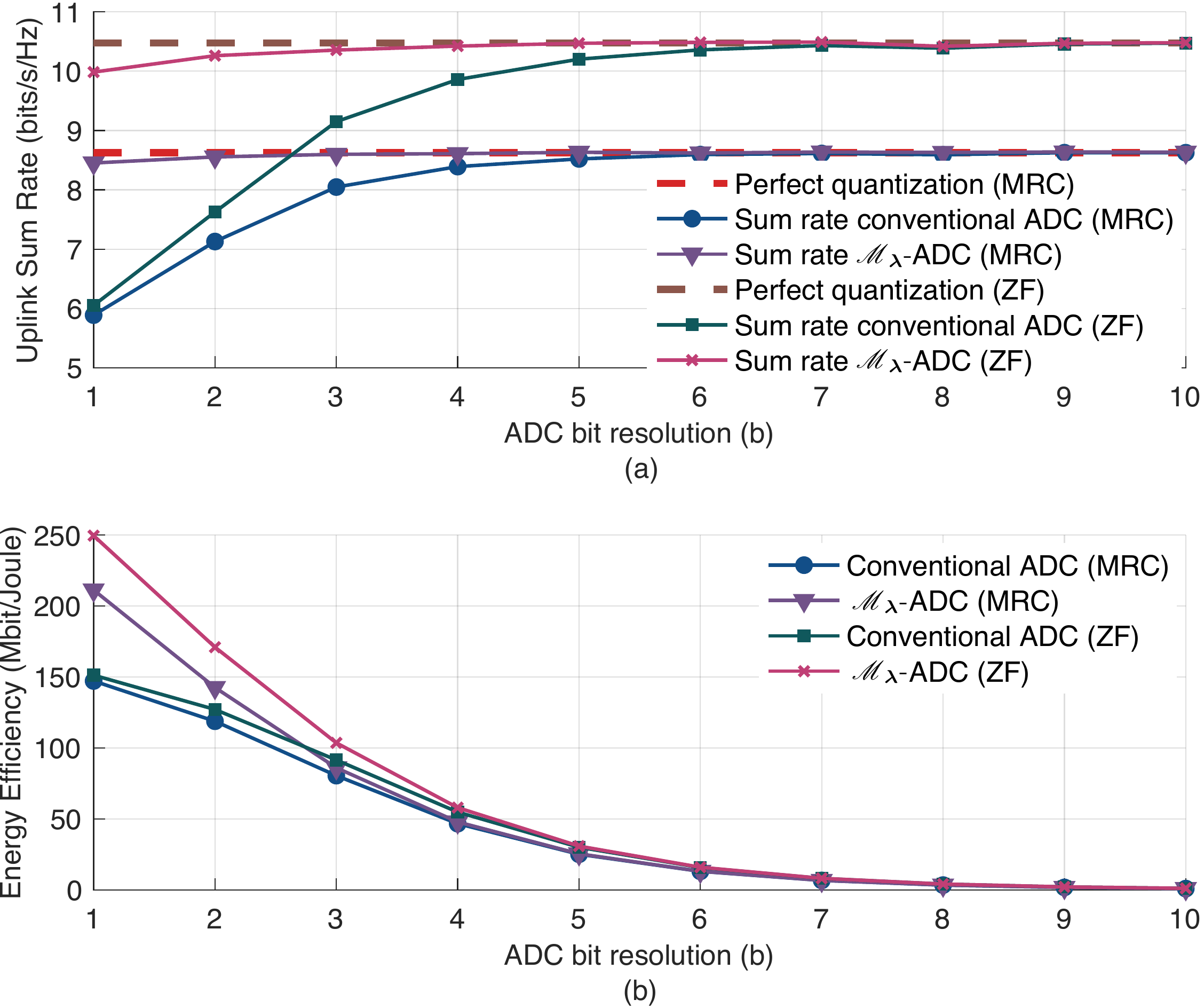}
		\caption{(a) Uplink sum-rate versus ADC bit resolution $b$, (b) Energy efficiency versus ADC bit resolution $b$.}
		\label{fig:sumrateAndevsb}
	\end{center}
\end{figure}
In this simulation, a hexagonal cell with a radius of $1000$m and a central disk radius $d_\mathsf{min}=100$m is considered. The large scale attenuation $\eta_\mathsf{n}$ is modeled as $\eta_\mathsf{n}=z_\mathsf{n} \left(d_m/ d_\mathsf{min }\right)^{-v}$, where $z_\mathsf{n}$ is a log-norm random variable with standard deviation $\sigma_\mathsf{shadow}=8$dB, $v=3.8$ is the path loss exponent, $d_m$ is the distance between $m^\textrm{th}$ user and BS. We assume a narrowband system and $D=1$, therefore the small-scale fading coefficient $g_{{n,m}} \sim \mathcal{CN} (0, 1)$. Additionally, $N=10$ users are randomly distributed with a transmit power $p_\mathsf{u} = 10$dB. 

Then, the achievable sum-rates of \madc and conventional ADC are compared in \fig{fig:mrcAndzfsr}. In \fig{fig:mrcAndzfsr} (a), (c), analytical approximations and simulation results for \madc and conventional ADCs with MRC detector and bit resolution $b=1, 2, \infty$ are shown. Moreover, results for \madc and conventional ADCs with ZF detector and  $b=2, 3, \infty$ are presented in \fig{fig:mrcAndzfsr} (b), (d). In all cases, a clear observation can be found that by adopting \madc the achievable uplink sum-rate can approach the sum-rate achieved by $b=\infty$. Also, the analytical approximations and simulations of achievable sum-rate for \madc are in agreement.

The power scaling law is also examined by setting the user transmit power to $p_\mathsf{u}={E_\mathsf{u}}/{M}$. We consider a scenario that $N=10$ users in the cell, $E_\mathsf{u}=10$ dB, $b = 1, 2, \infty$, and MRC detector in the BS with different number of antennas. According to \fig{fig:sreumrc}, the results show that \madc has close achievable uplink sum-rate even with $b=1$ to the sum-rate of conventional ADC with perfect quantization. In addition, with the increase of the number of BS antenna, there is no obvious rise on the achievable sum-rate. Thus, there is no need to equip the BS with very large quantity of antennas regarding achievable sum-rate. If the \madc is used, there is no need to have very high-resolution ($8$-$12$ bits) ADCs.

The relationship between achievable uplink sum-rate and ADC bit resolution $b$ and energy efficiency are investigated. Simulation results of uplink sum-rate versus bit resolution are presented for $\mathscr{M}_{\lambda}$ and conventional ADC with MRC and ZF detector in \fig{fig:sumrateAndevsb} (a). The simulation settings are users, $N=10$, $p_\mathsf{u}=10$ dB, and BS antennas, $M=50$. The observation is that \madc outperforms conventional ADC with respect to achievable sum-rate with low resolution $b$. In the case of MRC detector, the achievable uplink sum-rate of \madc with only 2 bit resolution nearly approaches the limit with infinite resolution. The similar result is also obtained in the ZF detector case, with $b=5$ bit resolution, the \madc can achieve close result to the conventional ADC with perfect quantization.  

Finally, the energy efficiency $\xi \triangleq \Omega R/P$ \cite{bai2015energy} is measured, where $\Omega =1$ MHz is the bandwidth, $R$ is the sum-rate. $P = c_0M2^b+c_1$ with $c_0 = 10^{-4}$ Watt and $c_1 = 0.02$ Watt. We can know that the energy efficiency of \madc is higher than conventional ADC when $b \leq 6$, and lower resolutions $b$ have higher energy efficiency. Considering the results in \fig{fig:sumrateAndevsb} (b) that the achievable sum-rate saturates after some resolution $b$ in MRC and ZF, low resolution $b$ (e.g., $b = 1$-$3$) of \madc should be selected to reach high energy efficiency and reduce power consumption.

\section{Conclusion}
\label{sec:con}

Modern communication systems are pivoted on the \mmimo technology. The success of this technology is tied to the power consumption and receiver/ADC saturation problems. In this work, we have described a new receiver design that can tackle the two challenges simultaneously. Our approach is based on the use of \madcs that are at the heart of the Unlimited Sensing Framework. By introducing modulo non-linearity in the RF-chain, we are able to fold arbitrary high dynamic range signals into low dynamic range modulo samples. This folding effect offers higher quantization resolution, given a fixed bit budget. The distinguishing features of our work include, (a) leveraging higher signal-to-quantization noise ratio (SQNR), (b) approaching the detection and average uplink sum-rate performances comparable to an $\infty$-bit resolution, conventional ADC, with a very few bit \madc. This is the key to enabling higher order modulation schemes \eg $1024$ QAM, (c) higher power efficiency attributing to a superior trade-off between energy efficiency and bit budget. We validate our approach using computer simulations and \madc hardware experiments. 

Given the end-to-end nature of our work, it opens up new research questions in the areas of theory, algorithms and hardware design. 

\begin{enumerate}[leftmargin = *,label = $\bullet$]
\item \textbf{Theoretical front.} 
Using \madc in the RF-chain necessitates the development of new theoretical models. This is important for precise understanding of the fundamental limits of our work. For instance, currently we have assumed that modulo threshold is low enough to map a Gaussian source into its uniform counterpart. Mathematical analysis based on the theory of wrapped distributions will guide towards a clearer quantitative trade-off between quantization noise and modulo threshold. This is the key to leveraging higher order modulation schemes which was previously highly challenging, specially with low-resolution ADCs.

\item \textbf{Algorithmic front.} Practical deployment of modulo ADCs \cite{Bhandari:2021:J} may result in non-ideal measurements. Further to this, influence of noise in the RF-chain remains to be investigated. These aspects necessitate the design of advanced algorithms for reconstruction and detection, specially the ones that can operate with moderate levels of oversampling \cite{Bhandari:2021:J}.

\item \textbf{Hardware front.} There are steps to be taken before our proof-of-concept hardware experiments can be implemented in practice. The first step in this direction would be a tight integration of \madc in the RF-chain. This requires custom designed hardware with technical specifications matched to the existing communication systems.
\end{enumerate}



\begin{thebibliography}{10}
\providecommand{\url}[1]{#1}
\csname url@samestyle\endcsname
\providecommand{\newblock}{\relax}
\providecommand{\bibinfo}[2]{#2}
\providecommand{\BIBentrySTDinterwordspacing}{\spaceskip=0pt\relax}
\providecommand{\BIBentryALTinterwordstretchfactor}{4}
\providecommand{\BIBentryALTinterwordspacing}{\spaceskip=\fontdimen2\font plus
\BIBentryALTinterwordstretchfactor\fontdimen3\font minus
  \fontdimen4\font\relax}
\providecommand{\BIBforeignlanguage}[2]{{%
\expandafter\ifx\csname l@#1\endcsname\relax
\typeout{** WARNING: IEEEtran.bst: No hyphenation pattern has been}%
\typeout{** loaded for the language `#1'. Using the pattern for}%
\typeout{** the default language instead.}%
\else
\language=\csname l@#1\endcsname
\fi
#2}}
\providecommand{\BIBdecl}{\relax}
\BIBdecl

\bibitem{larsson2014massive}
E.~G. Larsson, O.~Edfors, F.~Tufvesson, and T.~L. Marzetta, ``Massive {MIMO}
  for next generation wireless systems,'' \emph{{IEEE} Commun. Mag.}, vol.~52,
  no.~2, pp. 186--195, 2014.

\bibitem{andrews2014will}
J.~G. Andrews, S.~Buzzi, W.~Choi, S.~V. Hanly, A.~Lozano, A.~C. Soong, and
  J.~C. Zhang, ``What will {5G} be?'' \emph{{IEEE} J. Sel. Areas Commun.},
  vol.~32, no.~6, pp. 1065--1082, 2014.

\bibitem{gao2016energy}
X.~Gao, L.~Dai, S.~Han, I.~Chih-Lin, and R.~W. Heath, ``Energy-efficient hybrid
  analog and digital precoding for mmwave {MIMO} systems with large antenna
  arrays,'' \emph{{IEEE} J. Sel. Areas Commun.}, vol.~34, no.~4, pp. 998--1009,
  2016.

\bibitem{zhang2016spectral}
J.~Zhang, L.~Dai, S.~Sun, and Z.~Wang, ``{On the spectral efficiency of massive
  MIMO systems with low-resolution ADCs},'' \emph{{IEEE} Commun. Lett.},
  vol.~20, no.~5, pp. 842--845, 2016.

\bibitem{zhang2018low}
J.~Zhang, L.~Dai, X.~Li, Y.~Liu, and L.~Hanzo, ``On low-resolution {ADCs} in
  practical {5G} millimeter-wave massive {MIMO} systems,'' \emph{{IEEE} Commun.
  Mag.}, vol.~56, no.~7, pp. 205--211, 2018.

\bibitem{shannon1984communication}
C.~E. Shannon, ``Communication in the presence of noise,'' \emph{Proc. {IEEE}},
  vol.~72, no.~9, pp. 1192--1201, 1984.

\bibitem{fan2015uplink}
L.~Fan, S.~Jin, C.-K. Wen, and H.~Zhang, ``Uplink achievable rate for massive
  {MIMO} systems with low-resolution {ADC},'' \emph{{IEEE} Commun. Lett.},
  vol.~19, no.~12, pp. 2186--2189, 2015.

\bibitem{jacobsson2017throughput}
S.~Jacobsson, G.~Durisi, M.~Coldrey, U.~Gustavsson, and C.~Studer,
  ``{Throughput analysis of massive MIMO uplink with low-resolution ADCs},''
  \emph{{IEEE} Trans. Wireless Commun.}, vol.~16, no.~6, pp. 4038--4051, 2017.

\bibitem{choi2020advanced}
J.~Choi, G.~Lee, A.~Alkhateeb, A.~Gatherer, N.~Al-Dhahir, and B.~L. Evans,
  ``{Advanced receiver architectures for millimeter-wave communications with
  low-resolution ADCs},'' \emph{{IEEE} Commun. Mag.}, vol.~58, no.~8, pp.
  42--48, 2020.

\bibitem{mollen2016uplink}
C.~Mollen, J.~Choi, E.~G. Larsson, and R.~W. Heath, ``{Uplink performance of
  wideband massive {MIMO} with one-bit {ADC}s},'' \emph{{IEEE} Trans. Wireless
  Commun.}, vol.~16, no.~1, pp. 87--100, 2016.

\bibitem{7247358}
S.~Jacobsson, G.~Durisi, M.~Coldrey, U.~Gustavsson, and C.~Studer, ``{One-bit
  massive MIMO: Channel estimation and high-order modulations},'' in \emph{IEEE
  Intl. Conf. on Communication Workshop (ICCW)}, Jun. 2015, pp. 1304--1309.

\bibitem{halsig2014information}
T.~Halsig, L.~Landau, and G.~Fettweis, ``Information rates for
  faster-than-{Nyquist} signaling with 1-bit quantization and oversampling at
  the receiver,'' in \emph{IEEE Vehicular Tech. Conf. (VTC).}, May 2014, pp.
  1--5.

\bibitem{sit2004micropower}
J.-J. Sit and R.~Sarpeshkar, ``{A micropower logarithmic A/D with offset and
  temperature compensation},'' \emph{{IEEE} J. Solid-State Circuits}, vol.~39,
  no.~2, pp. 308--319, 2004.

\bibitem{lee20092}
J.~Lee, J.~Kang, S.~Park, J.~S. Seo, J.~Anders, J.~Guilherme, and M.~P. Flynn,
  ``A 2.5 {mW} 80 {dB} {DR} 36 {dB} {SNDR} 22 {MS}/s logarithmic pipeline
  {ADC},'' \emph{{IEEE} J. Solid-State Circuits}, vol.~44, no.~10, pp.
  2755--2765, 2009.

\bibitem{Bhandari:2017:C}
\BIBentryALTinterwordspacing
A.~Bhandari, F.~Krahmer, and R.~Raskar, ``On unlimited sampling,'' in
  \emph{Intl. Conf. on Sampling Theory and Applications (SampTA)}, Jul. 2017.
\BIBentrySTDinterwordspacing

\bibitem{Bhandari:2020:Ja}
------, ``On unlimited sampling and reconstruction,'' \emph{{IEEE} Trans. Sig.
  Proc.}, vol.~69, pp. 3827--3839, Dec. 2020.

\bibitem{Bhandari:2020:Pata}
------, ``Methods and apparatus for modulo sampling and recovery,'' US Patent
  US10\,651\,865B2, May, 2020.

\bibitem{Bhandari:2018:Ca}
------, ``Unlimited sampling of sparse signals,'' in \emph{{IEEE} Intl. Conf.
  on Acoustics, Speech and Signal Processing (ICASSP)}, Apr. 2018.

\bibitem{Bhandari:2018:C}
------, ``Unlimited sampling of sparse sinusoidal mixtures,'' in \emph{{IEEE}
  Intl. Sym. on Information Theory ({ISIT})}, Jun. 2018.

\bibitem{Bhandari:2019:C}
A.~Bhandari and F.~Krahmer, ``On identifiability in unlimited sampling,'' in
  \emph{Intl. Conf. on Sampling Theory and Applications (SampTA)}, Jul. 2019.

\bibitem{Bhandari:2021:J}
A.~Bhandari, F.~Krahmer, and T.~Poskitt, ``Unlimited sampling from theory to
  practice: {Fourier}-{Prony} recovery and prototype {ADC},'' \emph{{IEEE}
  Trans. Sig. Proc.}, vol.~70, pp. 1131--1141, Sep. 2021.

\bibitem{Bhandari:2022:J}
A.~Bhandari, ``Back in the {US}-{SR}: {Unlimited} sampling and sparse
  super-resolution with its hardware validation,'' \emph{{IEEE} Signal Process.
  Lett.}, vol.~29, pp. 1047--1051, Mar. 2022.

\bibitem{Logan:1984}
\BIBentryALTinterwordspacing
B.~F. Logan, ``Signals designed for recovery after clipping--{II}. {Fourier}
  transform theory of recovery,'' \emph{{AT}{\&}T Bell Laboratories Technical
  Journal}, vol.~63, no.~2, pp. 287--306, Feb. 1984.
\BIBentrySTDinterwordspacing

\bibitem{Abel:1991}
\BIBentryALTinterwordspacing
J.~Abel and J.~Smith, ``Restoring a clipped signal,'' in \emph{{IEEE} Intl.
  Conf. on Acoustics, Speech and Sig. Proc. ({ICASSP})}, May 1991.
\BIBentrySTDinterwordspacing

\bibitem{Adler:2012}
\BIBentryALTinterwordspacing
A.~Adler, V.~Emiya, M.~G. Jafari, M.~Elad, R.~Gribonval, and M.~D. Plumbley,
  ``Audio inpainting,'' \emph{{IEEE} Trans. Acoust., Speech, Signal Process.},
  vol.~20, no.~3, pp. 922--932, Mar. 2012.
\BIBentrySTDinterwordspacing

\bibitem{Esqueda:2016}
\BIBentryALTinterwordspacing
F.~Esqueda, S.~Bilbao, and V.~Valimaki, ``Aliasing reduction in clipped
  signals,'' \emph{{IEEE} Trans. Sig. Proc.}, vol.~64, no.~20, pp. 5255--5267,
  Oct. 2016.
\BIBentrySTDinterwordspacing

\bibitem{Romanov:2019}
E.~Romanov and O.~Ordentlich, ``Above the {Nyquist} rate, modulo folding does
  not hurt,'' \emph{{IEEE} Signal Process. Lett.}, vol.~26, no.~8, pp.
  1167--1171, Aug. 2019.

\bibitem{Gong:2021:J}
Y.~Gong, L.~Gan, and H.~Liu, ``Multi-channel modulo samplers constructed from
  gaussian integers,'' \emph{{IEEE} Signal Process. Lett.}, vol.~28, pp.
  1828--1832, 2021.

\bibitem{Rudresh:2018:C}
\BIBentryALTinterwordspacing
S.~Rudresh, A.~Adiga, B.~A. Shenoy, and C.~S. Seelamantula, ``Wavelet-based
  reconstruction for unlimited sampling,'' in \emph{{IEEE} Intl. Conf. on
  Acoustics, Speech and Sig. Proc. ({ICASSP})}, Apr. 2018, pp. 4584--4588.
\BIBentrySTDinterwordspacing

\bibitem{Musa:2018:C}
O.~Musa, P.~Jung, and N.~Goertz, ``Generalized approximate message passing for
  unlimited sampling of sparse signals,'' in \emph{{IEEE} Global Conf. on Sig.
  and Info. Proc. ({GlobalSIP})}, Nov. 2018.

\bibitem{Ordentlich:2018:J}
O.~Ordentlich, G.~Tabak, P.~K. Hanumolu, A.~C. Singer, and G.~W. Wornell, ``A
  modulo-based architecture for analog-to-digital conversion,'' \emph{{IEEE} J.
  Sel. Topics Signal Process.}, pp. 1--1, 2018.

\bibitem{Bhandari:2020:C}
A.~Bhandari and F.~Krahmer, ``{HDR} imaging from quantization noise,'' in
  \emph{{IEEE} Intl. Conf. on Image Processing ({ICIP})}, Oct. 2020, pp.
  101--105.

\bibitem{FernandezMenduina:2021:J}
S.~Fernandez-Menduina, F.~Krahmer, G.~Leus, and A.~Bhandari, ``Computational
  array signal processing via modulo non-linearities,'' \emph{{IEEE} Trans.
  Sig. Proc.}, vol.~70, pp. 2168--2179, Jul. 2021.

\bibitem{Florescu:2022:J}
D.~Florescu, F.~Krahmer, and A.~Bhandari, ``The surprising benefits of
  hysteresis in unlimited sampling: {Theory}, algorithms and experiments,''
  \emph{{IEEE} Trans. Sig. Proc.}, vol.~70, pp. 616--630, 2022.

\bibitem{ordonez2021full}
L.~G. Ordoñez, P.~Ferrand, M.~Duarte, M.~Guillaud, and G.~Yang, ``On
  full-duplex radios with modulo-{ADC}s,'' \emph{IEEE Open J. of the Comm.
  Society}, vol.~2, pp. 1279--1297, 2021.

\bibitem{beaulieu2004parametric}
N.~C. Beaulieu and M.~O. Damen, ``{Parametric construction of Nyquist-I
  pulses},'' \emph{{IEEE} Trans. Commun.}, vol.~52, no.~12, pp. 2134--2142,
  2004.

\bibitem{tse2005fundamentals}
D.~Tse and P.~Viswanath, \emph{Fundamentals of wireless communication}.\hskip
  1em plus 0.5em minus 0.4em\relax Cambridge University Press, 2005.

\bibitem{van1976maximum}
W.~van Etten, ``Maximum likelihood receiver for multiple channel transmission
  systems,'' \emph{{IEEE} Trans. Commun.}, vol.~24, no.~2, pp. 276--283, 1976.

\bibitem{clerckx2013mimo}
B.~Clerckx and C.~Oestges, \emph{MIMO wireless networks: {Channels}, techniques
  and standards for multi-antenna, multi-user and multi-cell systems}.\hskip
  1em plus 0.5em minus 0.4em\relax Academic Press, 2013.

\bibitem{gersho2012vector}
A.~Gersho and R.~M. Gray, \emph{Vector quantization and signal
  compression}.\hskip 1em plus 0.5em minus 0.4em\relax Springer Science \&
  Business Media, 2012, vol. 159.

\bibitem{wang2002video}
Y.~Wang, J.~Ostermann, and Y.-Q. Zhang, \emph{Video processing and
  communications}.\hskip 1em plus 0.5em minus 0.4em\relax Prentice hall Upper
  Saddle River, NJ, 2002, vol.~1.

\bibitem{dardari2006joint}
D.~Dardari, ``{Joint clip and quantization effects characterization in OFDM
  receivers},'' \emph{{IEEE} Trans. Circuits Syst. {I}}, vol.~53, no.~8, pp.
  1741--1748, 2006.

\bibitem{bernhard2012analytical}
M.~Bernhard, D.~R{\"o}rich, T.~Handte, and J.~Speidel, ``{Analytical and
  numerical studies of quantization effects in coherent optical OFDM
  transmission with 100 Gbit/s and beyond},'' \emph{ITG-Fachtagung Photonische
  Netze}, pp. 34--40, 2012.

\bibitem{panter1951quantization}
P.~Panter and W.~Dite, ``Quantization distortion in pulse-count modulation with
  nonuniform spacing of levels,'' \emph{Proceedings of the IRE}, vol.~39,
  no.~1, pp. 44--48, 1951.

\bibitem{mardia2000directional}
K.~V. Mardia, P.~E. Jupp, and K.~Mardia, \emph{Directional statistics}.\hskip
  1em plus 0.5em minus 0.4em\relax Wiley Online Library, 2000, vol.~2.

\bibitem{mo2015capacity}
J.~Mo and R.~W. Heath, ``Capacity analysis of one-bit quantized {MIMO} systems
  with transmitter channel state information,'' \emph{{IEEE} Trans. Sig.
  Proc.}, vol.~63, no.~20, pp. 5498--5512, 2015.

\bibitem{orhan2015low}
O.~Orhan, E.~Erkip, and S.~Rangan, ``Low power analog-to-digital conversion in
  millimeter wave systems: Impact of resolution and bandwidth on performance,''
  in \emph{Proc. Inf. Theory Appl. Workshop (ITA)}, Feb. 2015, pp. 191--198.

\bibitem{qiao2016spectral}
D.~Qiao, W.~Tan, Y.~Zhao, C.-K. Wen, and S.~Jin, ``Spectral efficiency for
  massive {MIMO} zero-forcing receiver with low-resolution {ADC},'' in
  \emph{Proc. 8th Int. Conf. Wireless Commun. Signal Process. (WCSP)}, Oct.
  2016, pp. 1--6.

\bibitem{bai2015energy}
Q.~Bai and J.~A. Nossek, ``Energy efficiency maximization for {5G}
  multi-antenna receivers,'' \emph{Trans. Emerg. Telecommun. Technol.},
  vol.~26, no.~1, pp. 3--14, 2015.

\end{thebibliography}
\end{document}